\documentclass[journal,draftclsnofoot,onecolumn,12pt,comsoc, twoside,a4paper]{IEEEtran}

\input{packages_commands_theorems.tex}

\usepackage[noamssymbols]{newtxmath}
\usepackage{eucal}

\usepackage[capitalise]{cleveref}
\Crefname{proposition}{Prop.}{Props.}
\Crefname{corollary}{Cor.}{Cors.}
\Crefname{section}{Sec.}{Secs.}

\begin{document}

\title{How Much Can D2D Communication Reduce Content Delivery Latency in Fog Networks with Edge Caching?}

\author{Roy~Karasik,~\IEEEmembership{Student Member,~IEEE,}
        Osvaldo~Simeone,~\IEEEmembership{Fellow,~IEEE,}
        and~Shlomo~Shamai~(Shitz),~\IEEEmembership{Fellow,~IEEE}% <-this % stops a space
\thanks{R. Karasik and S. Shamai are with the Department of Electrical Engineering, Technion, Haifa
	32000, Israel (e-mail:roy@campus.technion.ac.il)}% <-this % stops a space
\thanks{O. Simeone is with the Centre for Telecommunications Research,
	Department of Informatics, King’s College London, London WC2R 2LS, U.K.
	(e-mail: osvaldo.simeone@kcl.ac.uk).}% <-this % stops a space
\thanks{This work has been supported by the European Research Council (ERC) under the European Union’s Horizon 2020 Research and Innovation Programme (Grant Agreement Nos. 694630 and 725731).}}

%\markboth{IEEE Transactions on Communications}%
%{Submitted paper}

\maketitle

\begin{abstract}
A Fog-Radio Access Network (F-RAN) is studied in which cache-enabled Edge Nodes (ENs) with dedicated fronthaul connections to the cloud aim at delivering contents to mobile users. Using an information-theoretic approach, this work tackles the problem of quantifying the potential latency reduction that can be obtained by enabling Device-to-Device (D2D) communication over out-of-band broadcast links. Following prior work, the Normalized Delivery Time (NDT) --- a metric that captures the high signal-to-noise ratio worst-case latency --- is adopted as the performance criterion of interest. Joint edge caching, downlink transmission, and D2D communication policies based on compress-and-forward are proposed that are shown to be information-theoretically optimal to within a constant multiplicative factor of two for all values of the problem parameters, and to achieve the minimum NDT for a number of special cases. The analysis provides insights on the role of D2D cooperation in improving the delivery latency.
\end{abstract}


\section{Introduction}
\begin{figure}[!t]
	\centering
		\input{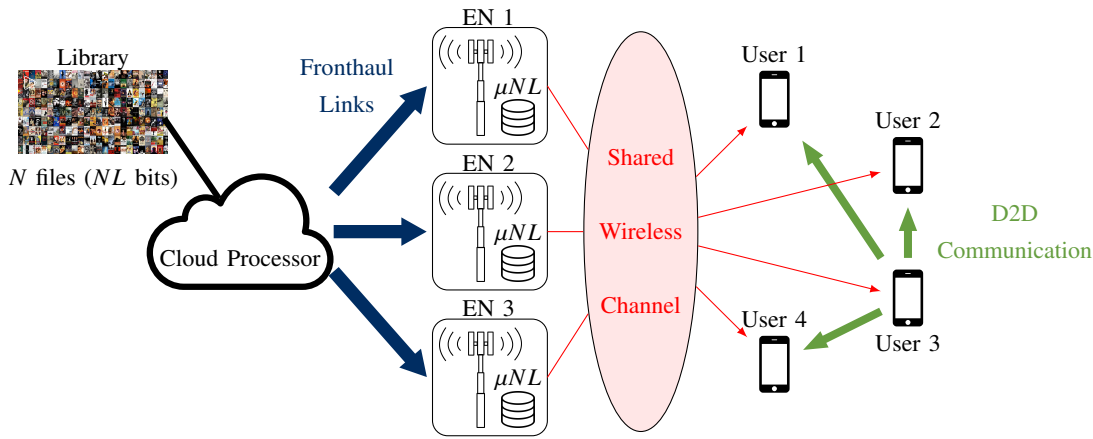}	
	\caption{Illustration of the D2D-aided F-RAN model under study with $M=3$ ENs and $K=4$ users.}
	\label{fig_model}
\end{figure}

\looseness=-1
\IEEEPARstart{P}{roactive} caching of popular content at the Edge Nodes (ENs) is an effective way of reducing delivery time \cite{li2018survey,parvez2018survey}. Apart from alleviating the need to access centralized network resources to fetch requested contents, edge caching also offers opportunities for cooperative transmission and interference management if there are common contents across the caches of multiple ENs. When requested contents are not cached at the edge, the ENs can satisfy the users' demands by leveraging fronthaul links to a Cloud Processor (CP) with full access to the content library. Fronthaul links can also enable cooperative transmission, as in a Cloud-Radio Access Network (C-RAN) architecture \cite{peng2014heterogeneous}. However, fronthaul transmissions generally entail additional latency. 
The Fog-RAN (F-RAN) architecture, illustrated in \cref{fig_model}, makes use of both cloud and edge caching resources in order to carry out content delivery, hence potentially reaping the benefits of both edge caching and C-RAN \cite{tandon2016harnessing,hung2015architecture,peng2016recent,chiang2016fog}.

Prior work, to be reviewed below, has studied the performance of F-RANs by assuming non-cooperative end users. In contrast, in this paper, motivated by the emergence of Device-to-Device (D2D) communication solutions \cite{asadi2014survey,bangerter2014networks,jameel2018survey}, we study the impact of D2D communication on the delivery latency of F-RAN architectures. To this end, we consider a D2D-aided F-RAN, illustrated in \cref{fig_model}, in which edge caching, fronthaul connectivity to a CP, and users' cooperation are leveraged to reduce content delivery time. 
We specifically aim at characterizing the potential latency reduction that may be achieved by utilizing out-of-band D2D links, while properly accounting for the latency overhead associated with D2D communications.

\looseness=-1
\textbf{Related Work:}
In prior work, the information-theoretic analysis of content delivery in F-RANs has been carried out in the high Signal-to-Noise-Ratio (SNR) regime in order to concentrate on the impact of interference. This line of work adopts as performance metric the Normalized Delivery Time (NDT), which measures the high-SNR worst-case latency relative to an ideal system with unlimited edge caching capability \cite{tandon2016harnessing,sengupta2016fog}. The first related work is \cite{maddah2015cache}, which presents an upper bound on the NDT, or equivalently, on the reciprocal Degrees-of-Freedom (DoF), for a cache-aided interference channel with three users.
Bounds on the NDT for arbitrary numbers of transmitters and receivers, where both transmitters and receivers have caching capabilities, were presented in \cite{hachem2016degrees} and in \cite{naderializadeh2017fundamental} under the constraint of linear precoders at the transmitters. A lower bound on the NDT was derived in \cite{sengupta2016cache} for any number of ENs and users, and it was shown to be tight for the setting of two ENs and two users. Upper and lower bounds on the NDT of a general interference channel with caches at all transmitters and receivers were presented in \cite{xu2017fundamental}, and the achievable NDT was shown to be optimal in certain cache size regimes.

\looseness=-1
Including also fronthaul connections to the cloud, the NDT of a general F-RAN system was investigated in \cite{sengupta2016fog}, where the proposed schemes were shown to achieve the minimum NDT to within a factor of 2, and the minimum NDT was completely characterized for two ENs and two users, as well as for other special cases. The F-RAN system with a shared multicast fronthaul link was studied in \cite{park2017coded} and \cite{kakar2018delivery}, where the advantages of coded multicast delivery were investigated. An F-RAN with heterogeneous contents was studied in \cite{goseling2017delivery}, and the NDT region was characterized for the case with two ENs and two users. A caching and delivery scheme was presented for a partially-connected F-RAN in \cite{roushdy2018cache} and in \cite{wan2018novel}. Under the constraints of linear precoding and uncoded fronthaul transmission, upper and lower bounds on the minimum NDT in an F-RAN were presented in \cite{zhang2017fundamental}, and the ratio between bounds was shown to be less than $3/2$ for all system parameters and equals to one for some special cases. This work was extended in \cite{shariatpanahi2019cloud} to include caches also at the users. An F-RAN with imperfect Channel State Information (CSI) at the CP was studied in \cite{zhang2018cloud}, and a non-orthogonal transmission scheme was shown to improve the latency performance. 

\looseness=-1
To the best of our knowledge, F-RANs with D2D communication have not yet been considered, apart from the conference versions of this work \cite{karasik2018ISIT,karasik2018iswcs}. Content delivery in a multi-hop D2D caching network was instead studied in \cite{jeon2017wireless}, where the per-node capacity scaling law was derived. In \cite{huang2009degrees}, it was shown that in-band transmitter or receiver cooperation cannot increase the sum DoF of an interference channel. In contrast, out-of-band D2D receiver cooperation was proven in \cite{wang2011interference} to increase the Generalized DoF metric for an interference channel. Importantly, reference \cite{wang2011interference} only imposes a rate constraint on the D2D links, hence not accounting for the latency overhead caused by D2D communications, which is of central interest in this work. The conference versions of this work cover the special case of an F-RAN with two ENs and users, whereas, in this work, as discussed next, we consider arbitrary numbers of ENs and users.

\looseness=-1
\textbf{Main Contributions:} 
In this work, we study the general D2D-aided F-RAN system with $M$ ENs and $K$ users illustrated in \cref{fig_model}. 
First, we propose two caching and delivery strategies based on a novel form of interference alignment and on compress-and-forward. The first strategy is developed for the special case $M=K=2$ and is shown to be optimal. The approach is however difficult to scale to a larger system and suffers from the typical lack of robustness to imperfect CSI of interference alignment \cite{motahari2014real}. For the general case of arbitrary number of $M$ and $K$, we prove that a more practical D2D strategy based on compress-and-forward achieves the minimum NDT to within a multiplicative factor of $2$. This implies that the optimality gap of this strategy does not scale with the size of the system. Based on these results, we identify regimes in terms of fronthaul and cache capacities under which D2D communication is beneficial in reducing delivery latency.

\textbf{Organization:} The rest of the paper is organized as follows. In Sec. \ref{sec:model}, we present an information-theoretic model for a general D2D-aided F-RAN under serial or pipelined delivery policies. In addition, the metric of interest, namely the NDT, is defined. In Sec. \ref{sec:x_channel}, we describe the proposed D2D-based caching and delivery strategies. In Sec. \ref{sec:bounds}, upper and lower bounds on the minimum NDT under serial delivery are derived. In Sec. \ref{sec:char}, we present an exact characterization of the minimum NDT for the special case with $M=K=2$ and a finite-gap characterization for arbitrary $M$ and $K$. In Sec. \ref{sec:pipelined}, we discuss pipelined delivery policies. Lower and upper bounds on the minimum NDT along with a finite-gap characterization are presented. Finally, in Sec. \ref{sec:conclusions} we conclude the paper and highlight some open problems.

\textbf{Notation:} For any positive integer $A$, we define the set $[A]\triangleq \{1,2,\ldots,A\}$. 

\section{System Model}\label{sec:model}
\looseness=-1
We consider the F-RAN system with Device-to-Device (D2D) links depicted in \cref{fig_model}, where $K\geq 2$ single-antenna users are served by $M\geq 2$ single-antenna Edge Nodes (ENs) over a downlink wireless channel. 
Each user is connected to all other users by an orthogonal out-of-band broadcast D2D link of capacity $C_D$ bits per symbol. The model generalizes the set-up studied in \cite{sengupta2016fog} by including D2D communications. Each EN is connected to a Cloud Processor (CP) by a fronthaul link of capacity $C_F$ bits per symbol. A symbol refers to a channel use of the downlink wireless channel.

Let $\mathcal F$ denote a library of $N\geq K$ files, $\mathcal F=\{f_1,\ldots,f_N\}$, each of size $L$ bits. The library is fixed for the considered time period. 
The entire library is available at the CP, whereas the ENs can only store up to $\mu NL$ bits each, where $0\leq\mu\leq 1$ is the fractional cache size. During the placement phase, contents are proactively cached at the ENs,
subject to the mentioned cache capacity constraints.

After the placement phase, the system enters the delivery phase, which is organized in Transmission Intervals (TIs). In every TI, each user arbitrarily requests one of the $N$ files from the library. The users' requests in a given TI are denoted by the demand vector $\mathbf d\triangleq (d_1,d_2,\ldots,d_K)\in[N]^K$. This vector is known at the beginning of a TI at the CP and ENs. The goal is to deliver the requested files to the users within the lowest possible delivery latency by leveraging fronthaul links, downlink channel, and D2D links.

For a given TI, let $T_E$ denote the duration of the transmission on the wireless downlink channel. At time $t\in[T_E]$, each user $k\in[K]$  receives a channel output given by 
\begin{IEEEeqnarray}{rCl}\label{eq:wireless_channel}
	y_k[t]&=&\sum_{m=1}^{M}h_{km}x_m[t]+z_k[t],
\end{IEEEeqnarray}
where $x_m[t]\in\mathbb C$ is the baseband symbol transmitted from EN $m\in[M]$ at time $t$, which is subject to the average power constraint $\mathbb E |x_m[t]|^2\leq P$ for some $P>0$; coefficient $h_{km}\in\mathbb C$ denotes the quasi-static flat-fading channel between EN $m$ to user $k$, which is assumed to be drawn independently and identically distributed (i.i.d.) from a continuous distribution and remain constant during each TI; and $z_k[t]$ is an additive white Gaussian noise, such that $z_k[t]\sim\mathcal C\mathcal N(0,1)$ is i.i.d. across time and users. The Channel State Information (CSI) $\mathbf{H}\triangleq\{h_{km}:k\in[K],m\in[M]\}$ is assumed to be known to all nodes.

\subsection{Caching, Delivery, and D2D Transmission}
\begin{figure}[!t]
	\centering
	\begin{tikzpicture}[>=latex]

\node (F) [thick,draw,minimum width=1.5cm,minimum height=0.6cm,fill=blue!60] at (0,0) {{\small Fronthaul}};
\node (E) [right = 0cm of F,thick,draw,minimum width=1.5cm,minimum height=0.6cm,fill=red!60] {{\small Edge}};
\node (D) [right = 0cm of E,thick,draw,minimum width=1.5cm,minimum height=0.6cm,fill=OliveGreen] {{\small D2D}};

\path[draw,<->] ($(F.north west)+(0,0.15cm)$) -- node[midway,above]{\small $T$} ($(D.north east)+(0,0.15cm)$);
\path[draw,<->] ($(F.south west)-(0,0.15cm)$) -- node[midway,below]{\small $T_F$} ($(F.south east)-(0,0.15cm)$);
\path[draw,<->] ($(E.south west)-(0,0.15cm)$) -- node[midway,below]{\small $T_E$} ($(E.south east)-(0,0.15cm)$);
\path[draw,<->] ($(D.south west)-(0,0.15cm)$) -- node[midway,below]{\small $T_D$} ($(D.south east)-(0,0.15cm)$);

\node (Ts) [below = 0.8cm of E] {{\footnotesize (a) Serial Transmission}};

\node (P) [right = 1cm of D,thick,draw,minimum width=4cm,minimum height=0.6cm,top color=blue!60,bottom color=OliveGreen, middle color=red!60]  {{\small Fronthaul + Edge + D2D}};

\path[draw,<->] ($(P.north west)+(0,0.15cm)$) -- node[midway,above]{\small $T$} ($(P.north east)+(0,0.15cm)$);

\node (Tp) [below = 0.8cm of P] {{\footnotesize (b) Pipelined Transmission}};

\end{tikzpicture}
	\caption{Transmission Interval structure for either serial or pipelined delivery policies.}
	\label{fig:TI}
\end{figure}
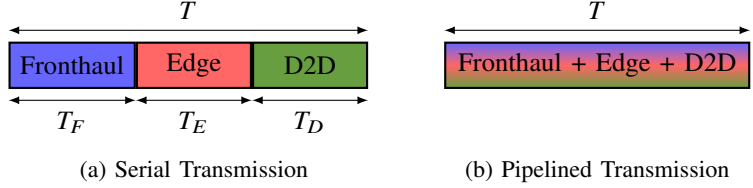

The operation of the system is defined by policies that perform caching, as well as delivery via fronthaul, edge, and D2D communication resources. For the delivery phase, there are generally two types of transmission policies, serial and pipelined. As detailed below, we first consider the serial transmission mode illustrated in \cref{fig:TI}a, and then, in \cref{sec:pipelined_model}, we describe the necessary adjustments to the delivery policies for allowing the pipelined simultaneous transmission mode illustrated in \cref{fig:TI}b.

\subsubsection{Caching Policy} During the placement phase, for EN $m$, $m\in[M]$, the caching policy is defined by functions $\pi^m_{c,n}(\cdot)$ that map each file $f_n$ to its cached content $s_{m,n}$ as
\begin{IEEEeqnarray}{rCl'l}\label{eq:caching_policy}
	s_{m,n}&=& \pi_{c,n}^m(f_n),&\forall n\in[N].
\end{IEEEeqnarray}
Note that, as per \eqref{eq:caching_policy}, we consider policies where only coding within each file is allowed, i.e., no inter-file coding (e.g., \cite{maddah-ali2014fundamental}) is permitted. In order to satisfy the cache capacity constraints, we restrict the mappings to satisfy $H(s_{m,n})\leq \mu L$. The overall cache content at EN $m$ is given by $s_m\triangleq(s_{m,1},s_{m,2}\ldots,s_{m,N})$. 
\subsubsection{Fronthaul Policy}In each TI of the delivery phase, for EN $m$, $m\in[M]$, the CP maps the library, $\mathcal F$, the demand vector $\mathbf{d}$ and CSI $\mathbf{H}$ to the fronthaul message
\begin{IEEEeqnarray}{rCl}
	\mathbf{u}_m=(u_m[1],u_m[2],\ldots,u_m[T_F])=\pi_f^m(\mathcal F,s_m,\mathbf{d},\mathbf{H}),
\end{IEEEeqnarray}
where $T_F$ is the duration of the fronthaul message. Note that the fronthaul message cannot exceed $T_FC_F$ bits, i.e., $H(\mathbf{u}_m)\leq T_FC_F$.
\subsubsection{Edge Transmission Policies}After fronthaul transmission, in each TI, the ENs transmit using a function $\pi_{e}^m(\cdot)$ that maps the local cache content, $s_m$, the received fronthaul message $\mathbf{u}_m$, the demand vector $\mathbf{d}$ and the global CSI $\mathbf{H}$, to the output codeword
\begin{IEEEeqnarray}{rClCl}
	\mathbf{x}_m&=&(x_m[1],x_m[2],\ldots,x_m[T_E])&=&\pi_{e}^m(s_m,\mathbf{u}_m,\mathbf{d},\mathbf{H}).
\end{IEEEeqnarray}
\subsubsection{D2D Interactive Communication Policies}
\looseness=-1
After receiving the signals \eqref{eq:wireless_channel} over $T_E$ symbols, in any TI, the users apply a D2D conferencing policy. For each user $k\in[K]$, this is defined by the interactive functions $\pi^k_{\text{D2D},t}(\cdot)$ that map the received signal $\mathbf{y}_k\triangleq (y_k[1],\ldots,y_k[T_E])$, the global CSI, and the previously received D2D message from users $[K]\backslash \{k\}$ to the D2D message
\begin{IEEEeqnarray}{l}\label{eq:d2d_policy}
	v_k[t]=
	\pi_{\text{D2D},t}^k\rb{\mathbf{y}_k,\mathbf{H},\mathbf{v}_{[K]}^{t-1}},
\end{IEEEeqnarray}
where $t\in[T_D]$, with $T_D$ being the duration of the D2D communication, and
\begin{IEEEeqnarray}{c}
	\mathbf{v}_{[K]}^{t-1}\triangleq \rb{v_1[1],\ldots,v_1[t-1],v_2[1],\ldots,v_2[t-1],\ldots,v_K[1],\ldots,v_K[t-1]}.
\end{IEEEeqnarray}
All users broadcast the D2D messages~\eqref{eq:d2d_policy} to all other users over orthogonal broadcast channels of capacity $C_D$. Hence, the total size of each D2D message cannot exceed $T_DC_D$ bits. i.e., $H(\mathbf{v}_k)\leq T_DC_D$, where $\mathbf{v}_k \triangleq (v_k[1],\ldots,v_k[T_D])$. 
 
\subsubsection{Decoding Policy}After D2D communication, each user $k\in[K]$ implements a decoding policy $\pi_d^k(\cdot)$ that maps the channel outputs, the D2D messages from users $[K]\backslash \{k\}$, the user demand, and the global CSI to an estimate of the requested file $f_{d_k}$ given as
\begin{IEEEeqnarray}{rCl}
	\hat{f}_{d_k}&=&\pi_d^k(\mathbf{y}_k,\mathcal{V}_{k},d_k,\mathbf{H}),
\end{IEEEeqnarray}
where $\mathcal{V}_{k}\triangleq \{\mathbf{v}_1,\ldots,\mathbf{v}_{k-1},\mathbf{v}_{k+1},\ldots,\mathbf{v}_K\}$
is the set of D2D messages sent by users $k'\in[K]\backslash \{k\}$ and received by user $k$.

The probability of error is defined as
\begin{IEEEeqnarray}{rCl}
	P_e&\triangleq&\max_{\mathbf{d}\in[N]^K}\max_{k\in[K]}\Pr(\hat{f}_{d_k}\neq f_{d_k}),
\end{IEEEeqnarray}
which is the worst-case probability of decoding error measured over all possible demand vectors $\mathbf{d}$ and over all users $k\in[K]$. A sequence of policies, indexed by the file size $L$, is said to be \emph{feasible} if, for almost all channel realization $\mathbf{H}$, we have $P_e\rightarrow0$ when $L\rightarrow\infty$.

\subsection{Performance Metric}
\looseness=-1
We adopt the Normalized Delivery Time (NDT), introduced in \cite{sengupta2016fog}, as the performance metric of interest. The NDT is the high-SNR ratio between the worst-case delivery time per bit required to satisfy any possible demand vector $\mathbf{d}$ and the delivery time per bit for an ideal reference system in which each user can receive the desired file at the maximum high-SNR rate of $\log(P)$ [bits/symbol].
To formalize the NDT, we parametrize fronthaul and D2D capacities as $C_F=r_F\log(P)$ and $C_D=r_D\log(P)$. With this parametrization, the fronthaul rate $r_F\geq 0$ represents the ratio between the fronthaul capacity and the high-SNR capacity of each EN-to-user wireless link in the absence of interference; a similar interpretation holds for the D2D rate $r_D\geq0$.

\looseness=-1
As discussed, under serial delivery, in each TI, the CP first sends the fronthaul messages to the ENs for a total time of $T_F$ symbols; then, the ENs transmit on the wireless shared channel for a total time of $T_E$ symbols; and, finally, the users use the out-of-band D2D links for a total time of $T_D$ symbols.
The corresponding NDT contributions are obtained by normalizing these terms by the delivery time needed on the mentioned reference system:
\begin{IEEEeqnarray}{c}
	\delta_F\triangleq\lim_{P\rightarrow\infty}\lim_{L\rightarrow\infty}\frac{\mathbb E[T_F]}{L/\log(P)},\quad \delta_E\triangleq\lim_{P\rightarrow\infty}\lim_{L\rightarrow\infty}\frac{\mathbb E[T_E]}{L/\log(P)},\quad \delta_D\triangleq\lim_{P\rightarrow\infty}\lim_{L\rightarrow\infty}\frac{\mathbb E[T_D]}{L/\log(P)}.\label{eq:F_NDT}\IEEEeqnarraynumspace
\end{IEEEeqnarray}
The factor $L/\log(P)$, used for normalizing the delivery times in~\eqref{eq:F_NDT}, represents the minimal time to deliver a file in the reference system. The total NDT under serial delivery is hence defined as
\begin{IEEEeqnarray}{rCl}\label{eq:NDT_def_S}
	\delta(\mu,r_F,r_D)&\triangleq&\delta_F+\delta_E+\delta_D,
\end{IEEEeqnarray}
where the notation emphasizes the dependence of the NDT on the fractional cache size $\mu$, and the fronthaul and D2D rates $r_F$ and $r_D$, respectively.

\looseness=-1
The minimum NDT is finally defined as the minimum over all NDTs achievable by some feasible policy:
\begin{IEEEeqnarray}{l}\label{eq:minimum_ndt_def}
	\delta^*(\mu,r_F,r_D)\triangleq
	\inf\{\delta(\mu,r_F,r_D):\delta(\mu,r_F,r_D)\text{ is achievable}\}.
\end{IEEEeqnarray} 
By construction, we have the lower bound $\delta^*(\mu,r_F,r_D)\geq 1$. Furthermore, the minimum NDT can be proved by means of file-splitting and cache-sharing arguments to be convex in $\mu$ for any fixed values of $r_F$ and $r_D$ \cite[Lemma 1]{sengupta2016fog}.

\subsection{Pipelined Transmission}\label{sec:pipelined_model}
The system defined above is based on serial delivery as illustrated in \cref{fig:TI}a.
Here we describe an alternative model, whereby, as seen in \cref{fig:TI}b, simultaneous transmissions on fronthaul, edge, and D2D channels are enabled. Specifically, the ENs can simultaneously receive messages over the fronthaul links and transmit on the wireless channel; and the users can receive on the wireless channel while, at the same time, transmitting messages on the D2D links. Following \cite{sengupta2016fog}, we refer to this model as enabling pipelined delivery.

\looseness=-1
To elaborate, at time instant $t\in[T]$, where $T$ denotes the delivery latency in a TI, each EN and user transmits using the information received at times $1,\ldots,t-1$, in a causal way. Mathematically, each EN $m\in[M]$ at time $t\in[T]$ uses a function $\pi_{\text{P},e,t}^m(\cdot)$ to map the local cache content, the fronthaul messages received up to time $t-1$, the demand vector, and the global CSI to the output symbol
\begin{IEEEeqnarray}{c}
	x_m[t]=\pi_{\text{P},e,t}^m(s_m,u_m[1],u_m[2],\ldots,u_m[t-1],\mathbf{d},\mathbf{H}).
\end{IEEEeqnarray}
Furthermore, user $k\in[K]$ transmits using the function $\pi^k_{\text{P},\text{D2D},t}(\cdot)$ that maps the received edge signal up to time $t-1$, global CSI, and the previously received D2D messages from users $[K]\backslash k$ to the D2D message
\begin{IEEEeqnarray}{c}
	v_k[t]=\pi_{\text{P},\text{D2D},t}^k\rb{y_k[1],\ldots,y_k[t-1],\mathbf{H},\mathbf{v}_{[K]}^{t-1}}.
\end{IEEEeqnarray}

Similar to the serial transmission case, the NDT and minimum NDT under pipelined delivery are defined as $\delta_\text{P}(\mu,r_F,r_D)\triangleq\lim_{P\rightarrow\infty}\lim_{L\rightarrow\infty}\mathbb E[T]\log(P)/L$, and $\delta_\text{P}^*(\mu,r_F,r_D)\triangleq
\inf\{\delta_\text{P}(\mu,r_F,r_D):\delta_\text{P}(\mu,r_F,r_D)\text{ is achievable}\}$,
respectively. Furthermore, we have the lower bound $\delta_\text{P}^*(\mu,r_F,r_D)\geq 1$, and the minimum NDT 
is a convex function of $\mu$ for any fixed values of $r_F$ and $r_D$. Finally, since serial delivery is a special case of pipelined delivery, by the definition of the minimum NDT, we have the inequality $\delta_\text{P}^*(\mu,r_F,r_D)\leq \delta^*(\mu,r_F,r_D)$. The pipelined delivery model is studied in Sec. \ref{sec:pipelined}.

\section{Delivery Strategies for Edge Caching with D2D Cooperation}\label{sec:x_channel}
In this section, we start by developing delivery schemes for the special case in which the fractional cache size is $\mu=1/M$ and the fronthaul capacity is not used. This scenario corresponds to the important special case in which the edge cache capacity is the minimum necessary to guarantee that the entire library $\mathcal F$ is available across the caches of all ENs, and hence fronthaul resources may not be used for delivery. 
Note that, for any request vector, users need to download equal fractions of the requested file from all ENs. This set-up is also known as an X-channel \cite{motahari2014real}.
We first introduce a delivery strategy based on a new interference alignment scheme for an F-RAN with $M=K=2$. A more scalable strategy based on compress-and-forward is then introduced for any number of ENs and users.

\subsection{Interference Alignment for $M=K=2$}\label{sec:2X2_ach}
For the case of $M=2$ ENs and $K=2$ users, we present a delivery scheme that integrates D2D communication in the Real Interference Alignment (RIA) scheme introduced in \cite{motahari2014real}. Our main interest in this scheme stems from its optimality, which will be proved in Sec. \ref{sec:char}.
\begin{proposition}\label{prop:ub_sic}
	For a D2D-aided F-RAN with $M=2$ ENs, each with a fractional cache size $\mu=1/2$, $K=2$ users, a fronthaul rate $r_F\geq 0$, and a D2D rate $r_D\geq 0$, the minimum NDT under serial delivery is upper bounded as $\delta^*(\mu=1/M,r_F,r_D)\leq \delta_\text{D2D-RIA}$, where 
	\begin{IEEEeqnarray}{c}\label{eq:ub_sic}
		\delta_\text{D2D-RIA}\triangleq 1+\frac{1}{2r_D}.
	\end{IEEEeqnarray}
\end{proposition}

\cref{prop:ub_sic} was proved in the conference paper \cite{karasik2018ISIT} by the authors by leveraging layered transmission, RIA, D2D cooperation, and successive cancellation decoding at the receivers. While referring to \cite{karasik2018ISIT} for details, we sketch here the main features of the scheme by comparing it to the original RIA scheme introduced in \cite{motahari2014real} for an X-channel model without D2D cooperation. In RIA, each EN applies layered transmission with two layers by transmitting
\begin{IEEEeqnarray}{c}\label{eq:ria_tx}
	x_1 = h_{22}a_1+h_{12}a_2\quad\text{and}\quad x_2=h_{21}b_1+h_{11}b_2,
\end{IEEEeqnarray}
where symbols $a_1$, $a_2$, $b_1$, and $b_2$ are chosen from a discrete constellation. Each layer is coded using random coding with rate $R$. Layers $a_1$ and $b_1$ are intended for user 1, whereas $a_2$ and $b_2$ are intended for user 2. Note that the precoders in~\eqref{eq:ria_tx} are based on perfect knowledge of the CSI at the ENs. The signals~\eqref{eq:wireless_channel} received by the two users are hence given as
\begin{IEEEeqnarray}{c}
	y_1=h_{11}h_{22}a_1+h_{12}h_{21}b_1+h_{11}h_{12}(a_2+b_2)+z_1,\IEEEnonumber\\
	y_2=h_{11}h_{22}b_2+h_{12}h_{21}a_2+h_{21}h_{22}(a_1+b_1)+z_2.
\end{IEEEeqnarray}
As shown in \cite{motahari2014real}, user 1 is able to decode the signal $\tilde{y}_1\triangleq y_1-z_1$ from $y_1$, in the high-SNR regime, if the rate is selected as $R=\log(P)/3$. Next, user 1, which has perfect CSI, searches for a set of symbols $\{a_1,b_1,a_2+b_2\}$ that generates $\tilde{y}_1$. Since the ENs use a discrete constellation and the channel coefficients are drawn i.i.d. from a continuous distribution, almost surely, this set is unique. This implies that user 1 can decode the desired layers $a_1$ and $b_1$ once it has decoded $\tilde{y}_1$. Similarly, user 2 can decode layers $a_2$ and $b_2$.
Note that the RIA scheme requires $T_E=3L/(2\log(P))$ channel uses in order to satisfy the users' demands, since each layer consists of $L/2$ bits and is transmitted at a rate of $\log(P)/3$ bits per channel use. It follows that RIA without D2D cooperation achieves an NDT of $3/2$.

In order to leverage D2D cooperation, in the proposed scheme, the ENs apply layered transmission with $n_d$ layers, where $n_d$ is odd. The transmitted signals are hence given as
\begin{IEEEeqnarray}{c}\label{eq:layers}
	x_1=\sum_{i=1}^{n_d} g_{1,i}a_i\quad\text{and}\quad x_2=\sum_{i=1}^{n_d} g_{2,i}b_i,
\end{IEEEeqnarray}
where precoder gains $\{g_{m,i}\}$, with $m\in[2]$ and $i\in[n_d]$, are selected to satisfy $h_{11}g_{1,i}=h_{12}g_{2,i-1}$ and $h_{22}g_{2,i}=h_{21}g_{1,i-1}$. The signals~\eqref{eq:wireless_channel} received by the two users are hence given as 
\begin{IEEEeqnarray}{rCl}
	y_1&=&h_{11}g_{1,1}a_1+\sum_{i=2}^{n_d}h_{11}g_{1,i}\rb{a_i+b_{i-1}}+h_{12}g_{2,n_d}b_{n_d}+z_1,\IEEEnonumber\\
	y_2&=&h_{22}g_{2,1}b_1+\sum_{i=2}^{n_d}h_{22}g_{2,i}\rb{b_i+a_{i-1}}+h_{21}g_{1,n_d}a_{n_d}+z_2.
\end{IEEEeqnarray}
In a manner similar to the RIA scheme, it can be shown that user 1 is able to decode the signal $\tilde{y}_1=y_1-z_1$ from $y_1$, in the high-SNR regime, if each layer is coded with rate $R=\log(P)/(n_d+1)$. Then, user 1 searches for the unique set $\mathcal {R}_1\triangleq\{a_1,a_2+b_1,\ldots,a_{n_d}+b_{n_d-1},b_{n_d}\}$ of symbols that generates $\tilde{y}_1$. The uniqueness of this set is determined by the same arguments used for the RIA scheme. Likewise, user 2 is able to identify the set $\mathcal {R}_2\triangleq\{b_1,b_2+a_1,\ldots,b_{n_d}+a_{n_d-1},a_{n_d}\}$.

\looseness=-1
In order to decode the desired layers, the users exchange the even-numbered layers over the D2D links, so that user 1 transmits the message $v_1=\{a_2+b_1,a_4+b_3,\ldots,a_{n_d-1}+b_{n_d-2}\}$ to user 2, whereas user 2 transmits $v_2=\{b_2+a_1,b_4+a_3,\ldots,b_{n_d-1}+a_{n_d-2}\}$ to user 1. User 1 is thus able to decode $\{a_1,b_2,a_3,b_4,\ldots,a_{n_d},b_{n_d}\}$ by means of successive cancellation decoding from $\{\mathcal{R}_1,v_2\}$. To this end, it starts by decoding $a_1$ in $\mathcal{R}_1$; then, it uses $a_1$ together with $b_2+a_1$ in $v_2$ to decode $b_2$; next, it uses $b_2$ and $a_3+b_2$ in $\mathcal{R}_1$ to decode $a_3$; and so on, until the desired layers are decoded. Similarly, user 2 decodes $\{b_1,a_2,b_3,a_4,\ldots,b_{n_d},a_{n_d}\}$ from $\{\mathcal{R}_2,v_1\}$. 

The scheme requires $T_E=(n_d+1)L/(n_d\log(P))$ downlink channel uses since each EN conveys $L/2$ bits to each user over $n_d/2$ layers, which are transmitted at a rate of $\log(P)/(n_d+1)$ bits per channel use. Unlike RIA, there is an additional latency overhead of $T_D=L/(2r_D\log(P))$ due to sharing $(n_d-1)/2$ layers over each D2D link. Therefore, assuming an arbitrarily large number of layers at the ENs, the NDT~\eqref{eq:ub_sic} is obtained. 

\subsection{Compress-and-Forward D2D Transmission}
\looseness=-1
The scheme discussed above appears to be cumbersome to generalize beyond the case $M=K=2$. Furthermore, at a practical level, this approach is mostly of theoretical interest since the performance of RIA is known to degrade catastrophically when CSI at the transmitters is imperfect \cite{davoodi2016aligned}.
Therefore, here we present an achievable scheme that applies to all values of $M$ and $K$ and requires only CSI at the receivers. The scheme is based on Compress-and-Forward (CF) D2D communication, and its near-optimality properties will also be discussed in Sec. \ref{sec:char}.

\looseness=-1
\begin{proposition}\label{prop:CF}
	For a D2D-aided F-RAN with $M$ ENs, each with a fractional cache size $\mu=1/M$, $K$ users, a library of $N\geq K$ files, a fronthaul rate $r_F\geq 0$, and a D2D rate $r_D\geq 0$, the minimum NDT under serial delivery is upper bounded as $\delta^*(\mu=1/M,r_F,r_D)\leq \delta_\text{D2D-CF}$, where the NDT
	\begin{IEEEeqnarray}{c}\label{eq:ach_ndt_cf}
		\delta_\text{D2D-CF}\triangleq \frac{K}{\min\{M,K\}}\rb{1+\frac{1}{r_D}}
	\end{IEEEeqnarray}
	is achieved by means of CF-based D2D communication and Zero-Forcing (ZF) equalization at the devices.
\end{proposition}

\looseness=-1
The NDT~\eqref{eq:ach_ndt_cf} is achieved by the following scheme. Consider first the case $M\geq K$. At any time, $K$ out of the $M$ ENs transmit simultaneously, each transmitting a fraction of the requested file to one of the $K$ users. As a result, the ENs' transmissions interfere at each user.
After downlink transmission, each user compresses and forwards its received signal to all other users over the D2D links. After D2D communication, each user collects the $K$ received signals, namely the signal that was directly received over the downlink channel and the compressed versions that were shared by the other users. Based on these signals, each user carries out ZF equalization in order to recover the desired signal with no interference from other signals. 

\looseness=-1
To elaborate, consider, for example, the case where the first $K$ ENs are active. After D2D cooperation, the signals $\mathbf{v}=[v_1,\ldots,v_K]^T$ available at user $k\in[K]$ can be expressed as $\mathbf{v}_k=\mathbf{H}_K\mathbf{x}+\mathbf{z}+\mathbf{q}_k$, where $\mathbf{x}\triangleq[x_1,\ldots,x_K]^T$ represents the transmitted signals, $\mathbf{H}_K$ is the channel matrix such that $(\mathbf{H}_K)_{i,j}=h_{ij}$, $\mathbf{z}\triangleq[z_1,\ldots,z_K]^T$ represents the white Gaussian noise, and $\mathbf{q}_k\triangleq[q_1,\ldots,q_K]^T$ represents the compression noise vector. We have $q_k=0$ since user $k$ receives $y_k$ directly over the downlink channel~\eqref{eq:wireless_channel}. The channel coefficients are drawn i.i.d. from a continuous distribution; therefore, almost surely, matrix $\mathbf{H}_K$ is invertible. Hence, each user can apply ZF equalization, i.e., multiply the received signals by $\mathbf{H}_K^{-1}$, to get $\mathbf{H}_K^{-1}\mathbf{v}_k=\mathbf{x}+\mathbf{H}_K^{-1}(\mathbf{z}+\mathbf{q}_k)$. Note that, after ZF equalization, the ENs' transmissions no longer cause interference. Therefore, the achievable rate is determined by the power of the additive noise $\mathbf{H}_K^{-1}(\mathbf{z}+\mathbf{q}_k)$. As shown in \cite[App. II-A]{sengupta2016fog}, by compressing with a rate equal to $\log(P)$ bits per downlink symbol, we can guarantee that the SNR after compression scales linearly with $P$. Thus, in the high-SNR regime, each EN is able to transmit with a rate of $R\approx\log(P)$ bits/channel use.

\looseness=-1
To satisfy the users' demands, each EN must convey $L/M$ bits to each user. To this end, we cluster the ENs into all possible $M\choose K$ subsets of $K$ ENs, and schedule each cluster into distinct time intervals of duration $T_E/{M\choose K}$. Since each EN participates in $M-1\choose K-1$ clusters, and the total number of bits transmitted by each EN is $KL/M$, then the duration of each interval is given as
\begin{IEEEeqnarray}{c}
	\frac{T_E}{{M\choose K}}=\frac{KL/M}{{M-1\choose K-1}R}=\frac{L}{{M\choose K}\log(P)}.
\end{IEEEeqnarray}
Therefore, the number of downlink channel uses is $T_E=L/\log(P)$, and hence the proposed scheme achieves an ideal edge NDT of $\delta_E=1$. Since, for each downlink channel use, each user transmits $\log(P)$ bits over the D2D link, a latency overhead of $T_D=T_E\log(P)/C_D=T_E/r_D$ is added to the delivery time, and hence the total NDT is~\eqref{eq:ach_ndt_cf}.

For the complementary case in which $M<K$, all ENs are active.
If $K$ is a multiple of $M$, then the users are partitioned into $K/M$ disjoint clusters of $M$ users. For each cluster, ZF equalization requires $L/\log(P)$ downlink channel uses in order to satisfy the demands of the users in the cluster. Therefore, the edge delivery time is $T_E=KL/(M\log(P))$, and hence the total NDT is~\eqref{eq:ach_ndt_cf}. For the more general case in which $K/M$ may not be an integer, the same edge delivery time can be achieved by clustering the users into all possible $K\choose M$ subsets of $M$ users, and, for each cluster, setting an interval of duration 
\begin{IEEEeqnarray}{c}
	\frac{T_E}{{K\choose M}}=\frac{L}{{K-1\choose M-1}\log(P)}.
\end{IEEEeqnarray}

\section{Bounds on the Minimum NDT for Serial Delivery}\label{sec:bounds}
In this section, we provide lower and upper bounds on the minimum NDT for the $M\times K$ D2D-aided F-RAN described in \cref{sec:model} in the case of serial delivery. 

\subsection{Upper Bounds and Achievable Strategy}\label{sec:serial_ub}
\looseness=-1
In the previous section, we presented schemes for the special case in which the fractional cache size is $\mu=1/M$. To obtain a policy that applies for any value of fractional cache size $\mu$, we combine, via file-splitting and cache-sharing, the D2D-based CF scheme (\cref{prop:CF}) with the best-known general strategies for an F-RAN model with no D2D cooperation. These strategies are described next for reference, followed by a review of file-splitting and cache-sharing.

\subsubsection{Cache-Aided ZF \cite[Lemma 2]{sengupta2016fog}} Cache-aided ZF precoding requires that all ENs cache the entire library of files, and hence it only applies for $\mu=1$. Full caching allows the ENs to cooperate by applying ZF-beamforming, whereby the precoding matrix equals the inverse of the channel matrix. This generates $\min\{M,K\}$ interference-free links to the users. Therefore, in the high-SNR regime, this scheme achieves a sum-rate of $\min\{M,K\}\log(P)$, and hence an NDT of $\delta_\text{ZF}\triangleq K/\min\{M,K\}$.

\subsubsection{Cache-Aided EN Coordination \cite[Lemma 3]{sengupta2016fog}} The RIA scheme discussed in Sec. \ref{sec:2X2_ach} can be applied to arbitrary number of ENs and users. Each EN transmits $M$ layers, and each layer is coded using random coding with rate $R\approx \log(P)/(M+K-1)$ bits per symbol. The layers are precoded such that, at each user, the desired layers can be decoded. The scheme hence achieves an NDT of $\delta_\text{IA}\triangleq (M+K-1)/M$.

\subsubsection{Cloud-Aided Soft-Transfer \cite[Proposition 3]{sengupta2016fog}} In this scheme, ZF precoding is carried out at the cloud, which has access to the entire library of files. The resulting encoded signals are then compressed with a resolution of $\log(P)$ bits per downlink baseband sample and conveyed to the ENs over the fronthaul links. Similar to the CF-based scheme (\cref{prop:CF}), it can be shown that the effective SNR in the downlink scales proportionally to the power $P$, and that this schemes achieves an NDT of $\delta_\text{ST}\triangleq K/\min\{M,K\}+K/(Mr_F)$, where the latency overhead of $\delta_F=K/(Mr_F)$ is due to transmission over the fronthaul links.

The described delivery techniques are combined by means of file-splitting and cache-sharing \cite[Lemma 1]{sengupta2016fog}. That is, all files are split in the same way into a number of fragments, and each fragment is delivered by using a different policy.
 
To formulate the main result, we define the threshold values
\begin{IEEEeqnarray}{c}\label{eq:D2D_th}
	r_F^\text{th}\triangleq\frac{K(M-1)}{M(\min\{M,K\}-1)}\quad\text{and}\quad r_D^\text{th}\triangleq \max\cb{\frac{\max\{M,K\}}{\min\{M,K\}-1},\frac{M^2r_F}{(M-1)\min\{M,K\}}}.\IEEEeqnarraynumspace
\end{IEEEeqnarray}

\begin{proposition}\label{prop:upper_bound}
	\looseness=-1
	For a D2D-aided F-RAN with $M$ ENs, each with a fractional cache size $\mu\in[0,1]$, $K$ users, a library of $N\geq K$ files, a fronthaul rate $r_F\geq 0$, and a D2D rate $r_D\geq 0$, the minimum NDT under serial delivery is upper bounded as $\delta^*(\mu,r_F,r_D)\leq\delta_\text{ach}(\mu,r_F,r_D)$, where the achievable NDT $\delta_\text{ach}(\mu,r_F,r_D)$ is obtained by combining the mentioned schemes as follows:
	\begin{itemize}
		\item Low cache, low fronthaul, and low D2D regime ($\mu\leq 1/M$, $r_F\leq r_F^\text{th}$, and $r_D\leq r_D^\text{th}$): Combining EN coordination and soft-transfer policies yields the NDT
		\begin{IEEEeqnarray}{c}
			\delta_\text{ach}(\mu,r_F,r_D)=(M+K-1)\mu+(1-\mu M)\sqb{\frac{K}{\min\{M,K\}}+\frac{K}{Mr_F}}.\label{eq:ach_low_low_low}
		\end{IEEEeqnarray}
		\item High cache, low fronthaul, and low D2D regime ($\mu> 1/M$, $r_F\leq r_F^\text{th}$, and $r_D\leq r_D^\text{th}$): Combining EN coordination and ZF precoding policies yields the NDT
		\begin{IEEEeqnarray}{c}
			\delta_\text{ach}(\mu,r_F,r_D)=\frac{K}{\min\{M,K\}}\rb{\frac{\mu M-1}{M-1}}+(1-\mu)\frac{M+K-1}{M-1}.\label{eq:ach_high_low_low}
		\end{IEEEeqnarray}
		\item High fronthaul and low D2D regime ($\mu\in[0,1]$, $r_F> r_F^\text{th}$, and $r_D\leq r_D^\text{th}$): Combining ZF precoding and soft-transfer policies yields the NDT
		\begin{IEEEeqnarray}{c}
			\delta_\text{ach}(\mu,r_F,r_D)=\frac{K}{\min\{M,K\}}+\frac{(1-\mu)K}{Mr_F}.\label{eq:ach_dc_high_low}
		\end{IEEEeqnarray}
		\item Low cache and high D2D regime ($\mu\leq 1/M$, $r_F\geq 0$, and $r_D> r_D^\text{th}$): Combining soft-transfer and CF policies yields the NDT
		\begin{IEEEeqnarray}{c}
			\delta_\text{ach}(\mu,r_F,r_D)=\frac{K}{\min\{M,K\}}\rb{1+\frac{\mu M}{r_D}}+(1-\mu M)\cdot\frac{K}{Mr_F}.\label{eq:ach_low_dc_high}
		\end{IEEEeqnarray}
		\item High cache and high D2D regime ($\mu> 1/M$, $r_F\geq 0$, and $r_D> r_D^\text{th}$): Combining CF and ZF precoding policies yields the NDT
		\begin{IEEEeqnarray}{c}
			\delta_\text{ach}(\mu,r_F,r_D)=\frac{K}{\min\{M,K\}}\rb{1+\frac{(1-\mu)M}{(M-1)r_D}}.\label{eq:ach_high_dc_high}
		\end{IEEEeqnarray}
	\end{itemize}
\end{proposition}
\begin{IEEEproof}
	See Appendix \ref{app:proof_ub_serial}.
\end{IEEEproof}

For the special case of $M=2$ ENs and $K=2$ users, the following NDT is achieved by using the D2D-enhanced RIA scheme of  Prop. \ref{prop:ub_sic}.
\begin{proposition}\label{prop:ub_2x2}
	\looseness=-1
	For a $2\times 2$ D2D-aided F-RAN with a fractional cache size $\mu\in[0,1]$, a library of $N\geq 2$ files, a fronthaul rate $r_F\geq 0$, and a D2D rate $r_D\geq 0$, the minimum NDT under serial delivery is upper bounded as $\delta^*(\mu,r_F,r_D)\leq\delta_{2\times2}(\mu,r_F,r_D)$, where
	\begin{IEEEeqnarray}{c}\label{eq:minimum_NDT}
		\delta_{2\times2}(\mu,r_F,r_D)\triangleq \left\lbrace\begin{array}{ll}
		\max\cb{1+\mu+\frac{1-2\mu}{r_F},2-\mu}&\text{for }0\leq r_F,r_D\leq 1,\\
		1+\frac{1-\mu}{r_F}&\text{for } r_F\geq\max\cb{1,r_D},\\
		\max\cb{1+\frac{\mu}{r_D}+\frac{1-2\mu}{r_F},1+\frac{1-\mu}{r_D}}&\text{for } r_D>\max\cb{1,r_F}.
		\end{array}\right.
	\end{IEEEeqnarray}
\end{proposition}
\begin{IEEEproof}
	Follows from \cref{prop:upper_bound} by replacing the D2D threshold $r_D^\text{th}$ in~\eqref{eq:D2D_th} with $r_D^\text{th}=\max\{1,r_F\}$, and, for D2D rate $r_D>r_D^\text{th}$, by applying the D2D scheme of \cref{prop:ub_sic} instead of the CF-based scheme.
\end{IEEEproof}

\subsection{Lower Bound}
A general lower bound on the minimum NDT is given in \cref{prop:lb}. Following \cite{sengupta2016fog}, the bound is derived by identifying subsets of information resources from which, for high-SNR, all requested files must be reliably decoded when a feasible policy is implemented. Specifically, for $l=0,1,\ldots,\min\{M,K\}$, we consider a subset that consists of the signals $\{\mathbf{y}_1,\ldots,\mathbf{y}_l,\mathcal{V}_1,\ldots,\mathcal{V}_l\}$ received by $l$ users on the downlink and D2D channels, along with the cache contents and fronthaul messages $\{s_1,\ldots,s_{(M-l)},\mathbf{u}_1,\ldots,\mathbf{u}_{(M-l)}\}$ of $(M-l)$ ENs.

\begin{proposition}\label{prop:lb}
	For a D2D-aided F-RAN with $M$ ENs, each with a fractional cache size $\mu\in[0,1]$, $K$ users, a library of $N\geq K$ files, a fronthaul rate $r_F\geq 0$, and a D2D rate $r_D\geq 0$, the minimum NDT under serial delivery is lower bounded as $\delta^*(\mu,r_F,r_D)\geq\delta_\text{lb}(\mu,r_F,r_D)$,	
	with $\delta_\text{lb}(\mu,r_F,r_D)$ being the minimum value of the following linear program
	\begin{subequations}\label{eq:linear_program}
		\begin{alignat}{2}
		&\!\text{minimize}&\qquad& \delta_F+\delta_E+\delta_D\\
		&\text{subject to}
		&& l\delta_E+(M-l)r_F\delta_F+g(l)r_D\delta_D\geq K-(M-l)(K-l)\mu,\label{eq:lb_family}\\
		&&& \delta_E\geq\frac{K}{\min\{M,K\}},\label{eq:lb_DoF}\\
		&&& \delta_F\geq 0,\;\delta_D\geq 0,\label{eq:lb_trivial}
		\end{alignat}
	\end{subequations}
\end{proposition}
where \eqref{eq:lb_family} is a family of constraints with $l=0,1,\ldots,\min\{M,K\}$, and
\begin{IEEEeqnarray}{c}\label{eq:g_l}
	g(l)\triangleq
	\begin{cases}
		0&\text{for }l=0,\\
		K-1&\text{for }l=1,\\
		K&\text{for }l=2,\ldots,\min\{M,K\}.
	\end{cases}
\end{IEEEeqnarray}
\begin{IEEEproof}
	See Appendix \ref{app:proof_lb}.
\end{IEEEproof}
\looseness=-1
Note that, without D2D communication, i.e., $r_D=0$, the linear program~\eqref{eq:linear_program} is identical to that of \cite[Proposition 1]{sengupta2016fog}. For $r_D>0$, the additional term $g(l)r_D\delta_D$ in~\eqref{eq:lb_family} reflects the novel trade-off between the D2D NDT $\delta_D$ and the edge and fronthaul NDTs $\delta_E$ and $\delta_F$, respectively. 

\section{Characterization of the Minimum NDT for Serial Delivery}\label{sec:char}
In this section, based on the lower and upper bounds presented in \cref{sec:bounds}, we discuss the optimality properties of the D2D-based strategies.
\subsection{$2\times 2$ D2D-Aided F-RAN}
For the case of $M=2$ ENs and $K=2$ users, as detailed in the following proposition, the D2D-based strategy of \cref{prop:ub_2x2} is optimal.
\begin{proposition}\label{prop:min_ndt_serial_2x2}
	The minimum NDT for the $2\times 2$ F-RAN system with number of files $N\geq 2$, a fractional cache size $\mu\in[0,1]$, a fronthaul rate $r_F\geq 0$, and a D2D rate $r_D\geq 0$ is given as $\delta^*\rb{\mu,r_F,r_D}=\delta_{2\times2}\rb{\mu,r_F,r_D}$.
\end{proposition}
\begin{IEEEproof}
	See Appendix \ref{app:proof_min_ndt_serial_2x2}.
\end{IEEEproof}

\looseness=-1
\cref{prop:min_ndt_serial_2x2} can be used to draw conclusions on the role of D2D cooperation in improving the delivery latency. We start by observing that, for $r_D\leq\max\{1,r_F\}$, the minimum NDT $\delta_{2\times2}\rb{\mu,r_F,r_D}$~\eqref{eq:minimum_NDT} is identical to the minimum NDT without D2D links derived in \cite[Corollary 3]{sengupta2016fog}. Therefore, D2D communication provides a latency reduction only when we have $r_D>\max\{1,r_F\}$.

The minimum useful value $\max\{1,r_F\}$ for the D2D rate $r_D$ increases with fronthaul rate $r_F$. This demonstrates that there exists a trade-off between fronthaul and D2D resources for the purpose of interference management, although their role is not symmetric. The use of fronthaul links is in fact necessary to obtain a finite NDT when the library is not fully available at the ENs, i.e., when $\mu<1/2$. D2D links can instead only reduce the NDT in regimes where fronthaul and edge resources would already be sufficient for content delivery with a finite NDT. In particular, when $r_D>\max\{1,r_F\}$, D2D communication reduces the minimum NDT for all values $0<\mu<1$. Furthermore, when $\mu>1/2$, irrespective of the value of $r_F$, the minimum NDT is achieved by leveraging only edge caching and D2D links, without having to rely on fronthaul resources, thus reducing the traffic at the network infrastructure. 

\subsection{General D2D-Aided F-RAN}
\looseness=-1
For arbitrary number of ENs and users, we start with the main result in the following proposition, which shows that the achievable CF-based strategy of \cref{prop:upper_bound} is optimal to within a multiplicative factor of two.

\begin{proposition}\label{prop:char}
	\looseness=-1
	For a D2D-aided F-RAN with $M$ ENs, each with a fractional cache size $\mu\in[0,1]$, $K$ users, a library of $N\geq K$ files, a fronthaul rate $r_F\geq 0$, and a D2D rate $r_D\geq 0$, the strategy of \cref{prop:upper_bound} achieves the minimum NDT under serial delivery to within a factor of two, i.e.,
	\begin{IEEEeqnarray}{c}\label{eq:NDT_ratio}
		\frac{\delta_\text{ach}(\mu,r_F,r_D)}{\delta^*(\mu,r_F,r_D)}\leq 2.
	\end{IEEEeqnarray}
\end{proposition}
\begin{IEEEproof}
	See Appendix \ref{app:proof_ratio2}.
\end{IEEEproof}

\begin{figure}[!t]
	\centering
	\resizebox {0.5\textwidth} {!} {	
		\begin{tikzpicture}

\begin{axis}[%
width=4.521in,
height=3.0in,
at={(0.758in,0.481in)},
scale only axis,
xmin=2,
xmax=10,
xlabel style={font=\color{white!15!black}},
xlabel={{\Large $M=K$}},
ymin=1.1,
ymax=1.9,
ylabel style={font=\color{white!15!black}},
ylabel={{\Large NDT}},
axis background/.style={fill=white},
xmajorgrids,
ymajorgrids,
legend style={at={(0.97,0.5)}, anchor=east, legend cell align=left, align=left, draw=white!15!black}
]
\addplot [color=blue, line width=2.0pt, mark size=4.0pt, mark=o, mark options={solid, blue}]
  table[row sep=crcr]{%
2	1.4\\
3	1.26666666666667\\
4	1.2\\
5	1.192\\
6	1.2\\
7	1.19591836734694\\
8	1.2\\
9	1.19753086419753\\
10	1.2\\
};
\addlegendentry{{\Large $\delta_\mathrm{lb}(\mu,r_F,r_D)$}}

\addplot [color=red, line width=2.0pt, mark size=2.8pt, mark=square, mark options={solid, red}]
  table[row sep=crcr]{%
2	1.5\\
3	1.66666666666667\\
4	1.75\\
5	1.8\\
6	1.8\\
7	1.8\\
8	1.8\\
9	1.8\\
10	1.8\\
};
\addlegendentry{{\Large $\delta_\mathrm{ach}(\mu,r_F,r_D)$}}

\addplot [color=red, dashed, line width=2.0pt, mark size=2.8pt, mark=square, mark options={solid, red}]
  table[row sep=crcr]{%
2	1.5\\
3	1.66666666666667\\
4	1.75\\
5	1.8\\
6	1.83333333333333\\
7	1.85714285714286\\
8	1.875\\
9	1.88888888888889\\
10	1.9\\
};
\addlegendentry{{\Large $\delta_\mathrm{ach}(\mu,r_F,0)$}}

\addplot [color=black, line width=1.5pt]
  table[row sep=crcr]{%
2	1.8\\
3	1.8\\
4	1.8\\
5	1.8\\
6	1.8\\
7	1.8\\
8	1.8\\
9	1.8\\
10	1.8\\
};
\addlegendentry{{\Large $\delta_\mathrm{D2D-CF}$}}

\addplot [color=red, line width=2.0pt, draw=none, mark size=4.0pt, mark=x, mark options={solid, red}, only marks]
  table[row sep=crcr]{%
2	1.4\\
};
\addlegendentry{{\Large $\delta^*(\mu,r_F,r_D)$ for $M=K=2$}}

\end{axis}
\end{tikzpicture}%
	}
	\caption{Lower and upper bounds on the minimum NDT as a function of the number of ENs and users $M=K$ for $r_F=1$, $\mu=1/M$, and $r_D=1.25$ or $r_D=0$.}
	\label{fig:serial_ndt_vs_M}
\end{figure}
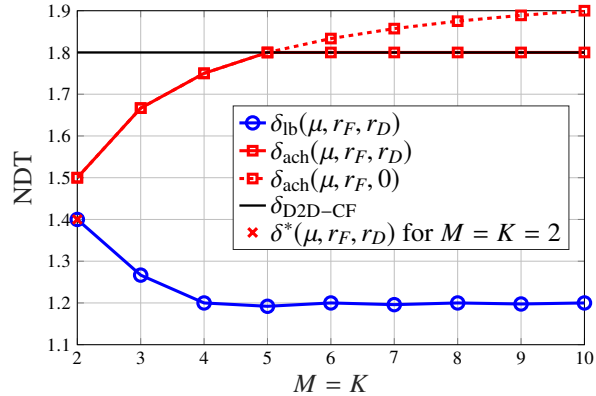

The key result in \cref{prop:char} is that the multiplicative suboptimality factor of the CF-based D2D approach defined in the previous section does not scale with the size of the system. This is illustrated in \cref{fig:serial_ndt_vs_M}, where we plot the achievable NDT $\delta_\text{ach}(\mu,r_F,r_D)$ and the lower bound $\delta_\text{lb}(\mu,r_F,r_D)$ as a function of the number of ENs and users, with $M=K$, fractional cache size $\mu=1/M$, fronthaul rate $r_F=1$, and D2D rate $r_D=1.25$. 
As seen, the suboptimality gap can be, in practice, significantly smaller than two.

While the gap identified in~\eqref{eq:NDT_ratio} is generally not zero, the next corollary states that CF is close to optimal for sufficiently high D2D rate.
\begin{corollary}\label{cor:aprox_optimal}
	For a D2D-aided F-RAN with $M$ ENs, each with a fractional cache size $\mu\in[0,1]$, $K$ users, a library of $N\geq K$ files, a fronthaul rate $r_F\geq 0$, and a D2D rate $r_D\geq\max\{r_D^\text{th},1/\epsilon\}$ with $r_D^\text{th}$ in~\eqref{eq:D2D_th} and $\epsilon>0$, the achievable strategy of \cref{prop:upper_bound} is close to optimal in the sense that we have
	\begin{IEEEeqnarray}{c}
		\frac{\delta_\text{ach}(\mu,r_F,r_D)}{\delta^*(\mu,r_F,r_D)}\leq 1+\epsilon.
	\end{IEEEeqnarray}
\end{corollary} 
\begin{IEEEproof}
	\cref{cor:aprox_optimal} follows directly from the proof of \cref{prop:char} (App. \ref{app:proof_ratio2}) since, for $r_D\geq r_D^\text{th}$, we have $\delta_\text{ach}(\mu,r_F,r_D)/\delta^*(\mu,r_F,r_D)\leq 1+1/r_D$ (cf.~\eqref{eq:1/rD_1} and \eqref{eq:1/rD_2}).
\end{IEEEproof}

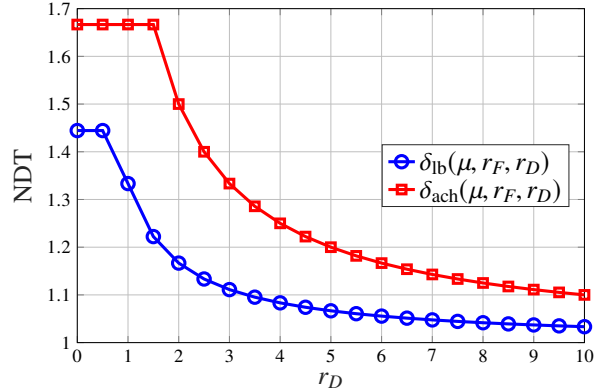
\begin{figure}[!t]
	\centering
	\resizebox {0.5\textwidth} {!} {	
		\begin{tikzpicture}

\begin{axis}[%
width=4.521in,
height=3in, 
at={(0.758in,0.481in)},
scale only axis,
xmin=0,
xmax=10,
xlabel style={font=\color{white!15!black}},
xlabel={{\Large $r_D$}},
ymin=1,
ymax=1.7,
ylabel style={font=\color{white!15!black}},
ylabel={{\Large NDT}},
axis background/.style={fill=white},
xmajorgrids,
ymajorgrids,
legend style={at={(0.97,0.5)}, anchor=east, legend cell align=left, align=left, draw=white!15!black}
]
\addplot [color=blue, line width=2.0pt, mark size=4.0pt, mark=o, mark options={solid, blue}]
  table[row sep=crcr]{%
0	1.44444444444444\\
0.5	1.44444444444444\\
1	1.33333333333333\\
1.5	1.22222222222222\\
2	1.16666666666667\\
2.5	1.13333333333333\\
3	1.11111111111111\\
3.5	1.0952380952381\\
4	1.08333333333333\\
4.5	1.07407407407407\\
5	1.06666666666667\\
5.5	1.06060606060606\\
6	1.05555555555556\\
6.5	1.05128205128205\\
7	1.04761904761905\\
7.5	1.04444444444444\\
8	1.04166666666667\\
8.5	1.03921568627451\\
9	1.03703703703704\\
9.5	1.03508771929825\\
10	1.03333333333333\\
};
\addlegendentry{{\Large $\delta_\mathrm{lb}(\mu,r_F,r_D)$}}

\addplot [color=red, line width=2.0pt, mark size=2.8pt, mark=square, mark options={solid, red}]
  table[row sep=crcr]{%
0	1.66666666666667\\
0.5	1.66666666666667\\
1	1.66666666666667\\
1.5	1.66666666666667\\
2	1.5\\
2.5	1.4\\
3	1.33333333333333\\
3.5	1.28571428571429\\
4	1.25\\
4.5	1.22222222222222\\
5	1.2\\
5.5	1.18181818181818\\
6	1.16666666666667\\
6.5	1.15384615384615\\
7	1.14285714285714\\
7.5	1.13333333333333\\
8	1.125\\
8.5	1.11764705882353\\
9	1.11111111111111\\
9.5	1.10526315789474\\
10	1.1\\
};
\addlegendentry{{\Large $\delta_\mathrm{ach}(\mu,r_F,r_D)$}}

\end{axis}
\end{tikzpicture}%
	}
	\caption{Lower and upper bounds on the minimum NDT as a function of $r_D$ for $r_F=1$, $M=K=3$, and $\mu=1/3$.}
	\label{fig:serial_ndt_vs_rD}
\end{figure}

\cref{cor:aprox_optimal} is illustrated in \cref{fig:serial_ndt_vs_rD}. where we plot the achievable NDT $\delta_\text{ach}(\mu,r_F,r_D)$ and the lower bound $\delta_\text{lb}(\mu,r_F,r_D)$ as a function of the D2D rate $r_D$, for $M=3$ ENs, $K=3$ users, fractional cache size $\mu=1/3$, and fronthaul rate $r_F=1$. 
As the D2D rate $r_D$ increases, the achievable NDT $\delta_\text{ach}(\mu,r_F,r_D)$ is seen to approach the lower bound $\delta_\text{lb}(\mu,r_F,r_D)$. For instance, for $r_D\geq 1/\epsilon=10$, the gap to optimality is smaller than $\epsilon=0.1$. This is because, for arbitrarily large D2D rate, the latency overhead caused by D2D communications is negligible, and an ideal NDT of one can be achieved by means of ZF-equalization at the users. 
In addition, the figure highlights the gains that can be achieved with sufficiently high D2D rate.

\section{Pipelined Delivery}\label{sec:pipelined}
\looseness=-1
In this section, we study the D2D-aided F-RAN model with pipelined delivery as defined in Sec. \ref{sec:pipelined_model}. We proceed in a manner similar to serial delivery by first deriving lower and upper bounds on the minimum NDT, and then discussing the optimality of CF-based D2D delivery. 

\subsection{Lower Bound on the Minimum NDT}
A lower bound on the minimum NDT for an $M\times K$ D2D-aided F-RAN under pipelined delivery policies is given in \cref{cor:piple_lb}. The lower bound is derived by following the same arguments as in \cref{prop:lb}, with the caveat that, under pipelined delivery policies, fronthaul, edge, and D2D transmissions occur simultaneously rather than sequentially.
\begin{corollary}\label{cor:piple_lb}
	\looseness=-1
	For a D2D-aided F-RAN with $M$ ENs, each with a fractional cache size $\mu\in[0,1]$, $K$ users, a library of $N\geq K$ files, a fronthaul rate $r_F\geq 0$, and a D2D rate $r_D\geq 0$, the minimum NDT under pipelined delivery is lower bounded as $\delta_\text{P}^*(\mu,r_F,r_D)\geq\delta_\text{P,lb}(\mu,r_F,r_D)$, where
	\begin{align}
	\delta_\text{P,lb}(\mu,r_F,r_D)=\max\left\lbrace\frac{K}{\min\{M,K\}},
	\max_{l=0,\ldots,\min\{M,K\}}\frac{K-(M-l)(K-l)\mu}{l+(M-l)r_F+g(l)r_D}\right\rbrace,
	\end{align}	
	and $g(l)$ is defined in~\eqref{eq:g_l}.
\end{corollary}
\begin{IEEEproof}
	Follows from the proof of \cref{prop:lb} (App. \ref{app:proof_lb}) with the following difference.
	For pipelined delivery policies, vectors $\mathbf{u}_m$, $\mathbf{x}_m$, $\mathbf{y}_k$, $\mathbf{z}_k$, and $\mathbf{v}_k$, which represents fronthaul message sent to EN $m\in[M]$, output codeword transmitted by EN $m$, signal received by user $k\in[K]$ on the shared wireless channel, white Gaussian noise at user $k$, and D2D message transmitted by user $k$, respectively, have $T$ entries, where $T$ is the delivery latency.
\end{IEEEproof}

\subsection{Upper Bound on the Minimum NDT}
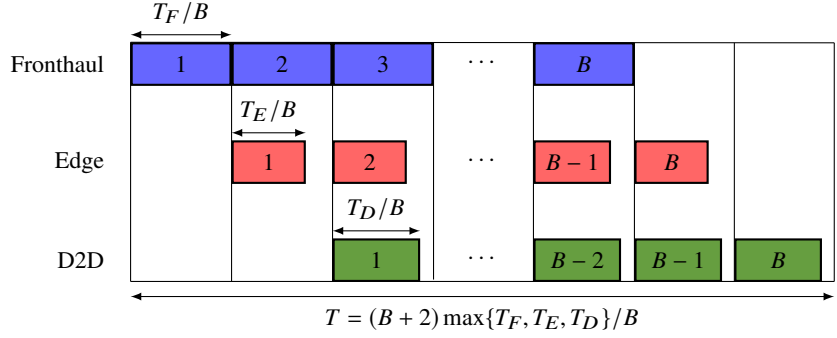
\begin{figure}[!t]
	\centering
		\begin{tikzpicture}[>=latex]
\node (TF) {{\footnotesize Fronthaul}};
\node (NF1) [right = 0.2cm of TF,thick,draw,minimum width={width("$B-1$")+10pt},minimum height=0.55cm,fill=blue!60] {{\footnotesize $1$}};
\node (NF2) [right = 0.0cm of NF1,thick,draw,minimum width={width("$B-1$")+10pt},minimum height=0.55cm,fill=blue!60] {{\footnotesize $2$}};
\node (NF3) [right = 0.0cm of NF2,thick,draw,minimum width={width("$B-1$")+10pt},minimum height=0.55cm,fill=blue!60] {{\footnotesize $3$}};
\node (NFdots) [right = 0.0cm of NF3,minimum width={width("$B-1$")+10pt},minimum height=0.55cm] {{\footnotesize $\cdots$}};
\node (NFB) [right = 0.0cm of NFdots,thick,draw,minimum width={width("$B-1$")+10pt},minimum height=0.55cm,fill=blue!60] {{\footnotesize $B$}};
\node (NFempty1) [right = 0.0cm of NFB,minimum width={width("$B-1$")+10pt},minimum height=0.55cm] {};
\node (NFempty2) [right = 0.0cm of NFempty1,minimum width={width("$B-1$")+10pt},minimum height=0.55cm] {};

\node (NEempty0) [below = 1.3cm of NF1.west,anchor=west,minimum width={width("$B-1$")+10pt},minimum height=0.55cm] {};
\node (NE1) [below = 1.3cm of NF2.west,anchor=west,thick,draw,minimum width={width("$B-1$")},minimum height=0.55cm,fill=red!60] {{\footnotesize $1$}};
\node (NE2) [below = 1.3cm of NF3.west,anchor=west,thick,draw,minimum width={width("$B-1$")},minimum height=0.55cm,fill=red!60] {{\footnotesize $2$}};
\node (NEdots) [below = 1.3cm of NFdots.west,anchor=west,minimum width={width("$B-1$")+10pt},minimum height=0.5cm] {{\footnotesize $\cdots$}};
\node (NEB-1) [below = 1.3cm of NFB.west,anchor=west,thick,draw,minimum width={width("$B-1$")},minimum height=0.55cm,fill=red!60] {{\footnotesize $B-1$}};
\node (NEB) [below = 1.3cm of NFempty1.west,anchor=west,thick,draw,minimum width={width("$B-1$")},minimum height=0.55cm,fill=red!60] {{\footnotesize $B$}};
\node (NEempty2) [below = 1.3cm of NFempty2.west,anchor=west,minimum width={width("$B-1$")+10pt},minimum height=0.55cm] {};
\node (TE) [left = 0.2cm of NEempty0] {{\footnotesize Edge}};

\node (NDempty0) [below = 1.3cm of NEempty0.west,anchor=west,minimum width={width("$B-1$")+10pt},minimum height=0.55cm] {};
\node (NDempty1) [below = 1.3cm of NE1.west,anchor=west,minimum width={width("$B-1$")+10pt},minimum height=0.55cm] {};
\node (ND1) [below = 1.3cm of NE2.west,anchor=west,thick,draw,minimum width={width("$B-1$")+5pt},minimum height=0.55cm,fill=OliveGreen] {{\footnotesize $1$}};
\node (NDdots) [below = 1.3cm of NEdots.west,anchor=west,minimum width={width("$B-1$")+10pt},minimum height=0.5cm] {{\footnotesize $\cdots$}};
\node (NDB-2) [below = 1.3cm of NEB-1.west,anchor=west,thick,draw,minimum width={width("$B-1$")+5pt},minimum height=0.55cm,fill=OliveGreen] {{\footnotesize $B-2$}};
\node (NDB-1) [below = 1.3cm of NEB.west,anchor=west,thick,draw,minimum width={width("$B-1$")+5pt},minimum height=0.55cm,fill=OliveGreen] {{\footnotesize $B-1$}};
\node (NDB) [below = 1.3cm of NEempty2.west,anchor=west,thick,draw,minimum width={width("$B-1$")+5pt},minimum height=0.55cm,fill=OliveGreen] {{\footnotesize $B$}};
\node (NDemptyX) [below = 1.3cm of NEempty2.west,anchor=west,minimum width={width("$B-1$")+10pt},minimum height=0.55cm] {};
\node (TD) [left = 0.2cm of NDempty0] {{\footnotesize D2D}};

\path[draw,<->] ($(NF1.north west)+(0,0.10cm)$) -- node[midway,above]{{\footnotesize $T_F/B$}} ($(NF1.north east)+(0,0.10cm)$);
\path[draw,<->] ($(NE1.north west)+(0,0.10cm)$) -- node[midway,above]{{\footnotesize $T_E/B$}} ($(NE1.north east)+(0,0.10cm)$);
\path[draw,<->] ($(ND1.north west)+(0,0.10cm)$) -- node[midway,above]{{\footnotesize $T_D/B$}} ($(ND1.north east)+(0,0.10cm)$);
\path[draw,<->] ($(NDempty0.south west)-(0,0.2cm)$) -- node[midway,below]{{\footnotesize $T=(B+2)\max\{T_F,T_E,T_D\}/B$}} ($(NDemptyX.south east)-(0,0.2cm)$);

\path[draw] (NF1.north west) -- (NDempty0.south west);
\path[draw] (NF2.north west) -- (NDempty1.south west);
\path[draw] (NF3.north west) -- (ND1.south west);
\path[draw] (NFdots.north west) -- (NDdots.south west);
\path[draw] (NFB.north west) -- (NDB-2.south west);
\path[draw] (NFempty1.north west) -- (NDB-1.south west);
\path[draw] (NFempty2.north west) -- (NDB.south west);
\path[draw] (NFempty2.north east) -- (NDemptyX.south east);
\path[draw] (NF1.north west) -- (NFempty2.north east);
\path[draw] (NDempty0.south west) -- (NDemptyX.south east);
\end{tikzpicture}
	\caption{Pipelining via block-Markov encoding.}
	\label{fig:block-Markov}
\end{figure}

\looseness=-1
To upper bound the minimum NDT, we consider a strategy that converts the CF-based serial transmission policies discussed in \cref{sec:serial_ub} into a pipelined delivery policy by means of block-Markov encoding and per-block file splitting. The approach is a generalization of the method presented in \cite[Sec. VI-B]{sengupta2016fog} for an F-RAN with no D2D links. To elaborate, fix a serial delivery policy with its fronthaul, edge, and D2D transmission strategy. As illustrated in Fig. \ref{fig:block-Markov}, in order to convert this strategy into one that leverages pipelining, every file in the library is split into $B$ blocks of size $L/B$ bits each, and every TI is divided into $B+2$ slots.
In each slot $b\in[B]$, the CP uses the fronthaul links to deliver the $b$th block of the requested files using the fronthaul transmission strategy of the selected serial policy. At the same time, the ENs, having received the fronthaul message for the $(b-1)$th block in the previous slot, apply the edge transmission strategy of the serial policy to deliver the $(b-1)$th block of the requested files to the users; and the users apply the corresponding conferencing scheme to cooperate in the decoding of the $(b-2)$th block of the requested files. 

For a serial delivery scheme that achieves fronthaul, edge, and D2D transmission durations $T_F$, $T_E$, and $T_D$, respectively, the block-Markov approach, with arbitrarily large number of blocks $B$, achieves the pipelined NDT 
\begin{IEEEeqnarray}{c}\label{eq:block-Markov}
	\delta_\text{P,ach}(\mu,r_F,r_D)=\lim_{B\rightarrow\infty}\lim_{P\rightarrow\infty}\lim_{L\rightarrow\infty}\frac{B+2}{B}\cdot\frac{\max\{T_F,T_E,T_D\}}{L/\log(P)}=\max\{\delta_F,\delta_E,\delta_D\},
\end{IEEEeqnarray}
where $\delta_F$, $\delta_E$, and $\delta_D$ are the fronthaul, edge, and D2D NDTs of the serial transmission scheme as defined in \eqref{eq:F_NDT}. 
Moreover, for two serial transmission schemes, one that achieves NDTs $\delta_F^{(1)}$, $\delta_E^{(1)}$, and $\delta_D^{(1)}$, whereas the other achieves NDTs $\delta_F^{(2)}$, $\delta_E^{(2)}$, and $\delta_D^{(2)}$, and for some $\alpha\in[0,1]$, the following pipelined NDT is achievable \cite[Sec. VI-B]{sengupta2016fog}
\begin{IEEEeqnarray}{c}
	\delta_\text{P,ach}(\mu,r_F,r_D)=
	\max\left\lbrace\alpha\delta_F^{(1)}+(1-\alpha)\delta_F^{(2)},\alpha\delta_E^{(1)}+(1-\alpha)\delta_E^{(2)},\alpha\delta_D^{(1)}+(1-\alpha)\delta_D^{(2)}\right\rbrace.\IEEEeqnarraynumspace
\end{IEEEeqnarray}
\begin{proposition}\label{prop:upper_bound_pipe}
	For an $M\times K$ D2D-aided F-RAN with a fractional cache size $\mu\in[0,1]$, a library of $N\geq K$ files, a fronthaul rate $r_F\geq 0$, and a D2D rate $r_D\geq 0$, the minimum NDT under pipelined delivery is upper bounded as
	$\delta^*(\mu,r_F,r_D)\leq\delta_\text{P,ach}(\mu,r_F,r_D)$, where the achievable NDT $\delta_\text{P,ach}(\mu,r_F,r_D)$ is given for two distinct regimes of operation as follows:
	\begin{itemize}
		\item High fronthaul rate ($r_F\geq \min\{M,K\}/M$):
		\begin{IEEEeqnarray}{c}\label{eq:pipe_ach_ndt1}
			\delta_\text{P,ach}(\mu,r_F,r_D)=\frac{K}{\min\{M,K\}}.
		\end{IEEEeqnarray}
		\item Low fronthaul rate ($r_F< \min\{M,K\}/M$):
		\begin{IEEEeqnarray}{c}\label{eq:pipe_ach_ndt2}
			\delta_\text{P,ach}(\mu,r_F,r_D)=\left\lbrace\begin{array}{ll}
				\frac{(1-M\mu)K}{Mr_F}&\text{for }\mu\in[0,\mu_1],\\
				\frac{(1-M\mu_1)K}{Mr_F}\cdot\frac{\mu_2-\mu}{\mu_2-\mu_1}
				+\frac{K}{\min\{M,K\}}\cdot\frac{\mu-\mu_1}{\mu_2-\mu_1}&\text{for }\mu\in(\mu_1,\mu_2),\\
				\frac{K}{\min\{M,K\}}&\text{for }\mu\in[\mu_2,1],
			\end{array}\right.
		\end{IEEEeqnarray}
		where we have defined
		\begin{IEEEeqnarray}{c}\label{eq:pipe_mu1}
			\mu_1\triangleq\frac{K-\max\{M,K\}r_F}{KM+Mr_F[\min\{M,K\}-1]},
		\end{IEEEeqnarray}
		and
		\begin{IEEEeqnarray}{c}\label{eq:pipe_mu2}
			\mu_2\triangleq \max\cb{1-\frac{Mr_F}{\min\{M,K\}}-\frac{M-1}{M}r_D,\frac{1}{M}-\frac{r_F}{\min\{M,K\}}}.
		\end{IEEEeqnarray}
	\end{itemize}
\end{proposition}
\begin{IEEEproof}
	See Appendix \ref{app:proof_ub_pipe}. 
\end{IEEEproof}

\subsection{Characterization of the Minimum NDT}
In the following propositions we discuss the optimality of the D2D CF-based strategy under pipelined delivery. First, we prove that the multiplicative suboptimality factor of two, identified in \cref{prop:char}, applies also to pipelined delivery policies. 
\begin{proposition}\label{prop:pipe_ndt_ratio}
	\looseness=-1
	For a D2D-aided F-RAN with $M$ ENs, $K$ users, a library of $N\geq K$ files, a fronthaul rate $r_F<\min\{M,K\}/M$, and a D2D rate $r_D< 1-Mr_F/\min\{M,K\}$, the strategy of \cref{prop:upper_bound_pipe} achieves the minimum NDT under pipelined delivery to within a factor of two, i.e.,
	\begin{IEEEeqnarray}{c}\label{eq:pipe_NDT_ratio}
		\frac{\delta_\text{P,ach}(\mu,r_F,r_D)}{\delta^*_\text{P}(\mu,r_F,r_D)}\leq 2,\quad\forall\mu\in[\mu_1,\mu_2]
	\end{IEEEeqnarray}
\end{proposition}
\begin{IEEEproof}
	See Appendix \ref{app:pipe_ndt_ratio}.
\end{IEEEproof}

Next, we show that the achievable strategy of \cref{prop:upper_bound_pipe} is optimal for the high fronthaul regime with $r_F\geq \min\{M,K\}/M$; for the high D2D regime with $r_D\geq 1-Mr_F/\min\{M,K\}$; for the low cache regime with $\mu\in[0,\mu_1]$; and for the high cache regime with $\mu\in[\mu_2,1]$.
\begin{proposition}\label{prop:pipe_min_NDT}
	For a D2D-aided F-RAN with $M$ ENs, each with a fractional cache size $\mu\in[0,1]$, $K$ users, a library of $N\geq K$ files, a fronthaul rate $r_F\geq 0$, and a D2D rate $r_D\geq 0$, the minimum NDT is characterized for three distinct regimes of operation as follows:
	\begin{itemize}
		\item High fronthaul rate ($r_F\geq \min\{M,K\}/M$):
		\begin{IEEEeqnarray}{c}\label{eq:pipe_min_NDT1}
			\delta_\text{P}^*(\mu,r_F,r_D)=\frac{K}{\min\{M,K\}}.
		\end{IEEEeqnarray}
		\item Low fronthaul rate and high D2D rate ($r_F< \min\{M,K\}/M$ and $r_D\geq 1-Mr_F/\min\{M,K\}$):
		\begin{IEEEeqnarray}{c}\label{eq:pipe_min_ndt_high_rd}
			\delta^*_\text{P}(\mu,r_F,r_D)=\max\cb{\frac{(1-M\mu)K}{Mr_F},\frac{K}{\min\{M,K\}}}.
		\end{IEEEeqnarray}
		\item Low fronthaul and D2D rates ($r_F< \min\{M,K\}/M$ and $r_D< 1-Mr_F/\min\{M,K\}$):
		\begin{IEEEeqnarray}{c}\label{eq:pipe_min_NDT2}
			\delta^*_\text{P}(\mu,r_F,r_D)=\left\lbrace\begin{array}{ll}
				\frac{(1-M\mu)K}{Mr_F}&\text{for }\mu\in[0,\mu_1],\\
				\frac{K}{\min\{M,K\}}&\text{for }\mu\in[\mu_2,1],
			\end{array}\right.
		\end{IEEEeqnarray}
		where $\mu_1$ and $\mu_2$ are defined in~\eqref{eq:pipe_mu1} and~\eqref{eq:pipe_mu2}, respectively.
	\end{itemize}
\end{proposition}
\begin{IEEEproof}
	See Appendix \ref{app:proof_pipe_min_ndt}.
\end{IEEEproof}

In the pipelined case, as seen in \cref{fig:block-Markov}, the latency is dictated by the largest among fronthaul, D2D, and edge NDTs. Therefore, whenever the fronthaul rate is large enough to enable ZF precoding on the wireless channel without causing a bottleneck, the minimum NDT can be achieved without using D2D communication. However, for low fronthaul rate and low cache capacity, cooperation via CF-based ZF equalization allows the delivery latency to be reduced by alleviating fronthaul load without increasing the edge NDT.

Comparing the results for serial and pipelined delivery policies, we observe that both the achievable NDT in \cref{prop:upper_bound} and the lower bound in \cref{prop:lb} are strictly decreasing functions of $r_D$ for all $r_D\geq r_D^\text{th}$, and hence the minimum NDT under serial delivery is strictly decreasing as well (cf. \cref{fig:serial_ndt_vs_rD}). In contrast, under pipelined delivery, the minimum NDT \eqref{eq:pipe_min_ndt_high_rd} for large $r_D$ is a constant function of $r_D$. This is because, when $r_D\geq 1-Mr_F/\min\{M,K\}$, the duration of the D2D transmission in each slot of the optimal block-Markov strategy is smaller than the fronthaul or edge transmissions, and hence increasing the D2D rate further does no reduce the minimum NDT.

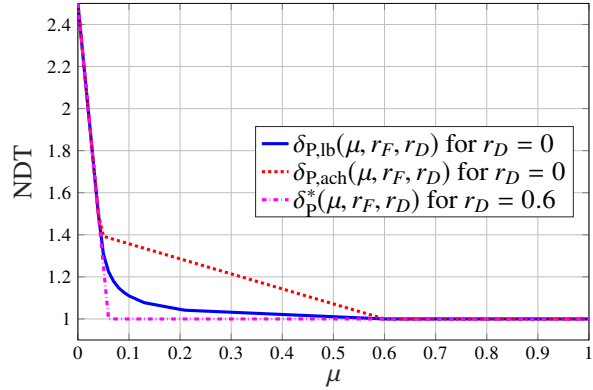
\begin{figure}[!t]
	\centering
	\resizebox {0.5\textwidth} {!} {	
		\definecolor{mycolor1}{rgb}{1.00000,0.00000,1.00000}%
\begin{tikzpicture}

\begin{axis}[%
width=4.521in,
height= 3in, 
at={(0.758in,0.481in)},
scale only axis,
xmin=0,
xmax=1,
xtick={  0, 0.1, 0.2, 0.3, 0.4, 0.5, 0.6, 0.7, 0.8, 0.9,   1},
xlabel style={font=\color{white!15!black}},
xlabel={{\Large $\mu$}},
ymin=0.9,
ymax=2.5,
ytick={  0, 0.2, 0.4, 0.6, 0.8,   1, 1.2, 1.4, 1.6, 1.8,   2, 2.2, 2.4},
ylabel style={font=\color{white!15!black}},
ylabel={{\Large NDT}},
axis background/.style={fill=white},
xmajorgrids,
ymajorgrids,
legend style={at={(0.97,0.5)}, anchor=east, legend cell align=left, align=left, draw=white!15!black}
]
\addplot [color=blue, line width=2.0pt]
  table[row sep=crcr]{%
0	2.5\\
0.01	2.25\\
0.02	2\\
0.03	1.75\\
0.04	1.5\\
0.05	1.30769230769231\\
0.06	1.225\\
0.07	1.17857142857143\\
0.08	1.14736842105263\\
0.09	1.12631578947368\\
0.1	1.10975609756098\\
0.11	1.09878048780488\\
0.12	1.08780487804878\\
0.13	1.07727272727273\\
0.14	1.07272727272727\\
0.15	1.06818181818182\\
0.16	1.06363636363636\\
0.17	1.05909090909091\\
0.18	1.05454545454545\\
0.19	1.05\\
0.2	1.04545454545455\\
0.21	1.04148936170213\\
0.22	1.04042553191489\\
0.23	1.03936170212766\\
0.24	1.03829787234043\\
0.25	1.03723404255319\\
0.26	1.03617021276596\\
0.27	1.03510638297872\\
0.28	1.03404255319149\\
0.29	1.03297872340426\\
0.3	1.03191489361702\\
0.31	1.03085106382979\\
0.32	1.02978723404255\\
0.33	1.02872340425532\\
0.34	1.02765957446808\\
0.35	1.02659574468085\\
0.36	1.02553191489362\\
0.37	1.02446808510638\\
0.38	1.02340425531915\\
0.39	1.02234042553191\\
0.4	1.02127659574468\\
0.41	1.02021276595745\\
0.42	1.01914893617021\\
0.43	1.01808510638298\\
0.44	1.01702127659574\\
0.45	1.01595744680851\\
0.46	1.01489361702128\\
0.47	1.01382978723404\\
0.48	1.01276595744681\\
0.49	1.01170212765957\\
0.5	1.01063829787234\\
0.51	1.00957446808511\\
0.52	1.00851063829787\\
0.53	1.00744680851064\\
0.54	1.0063829787234\\
0.55	1.00531914893617\\
0.56	1.00425531914894\\
0.57	1.0031914893617\\
0.58	1.00212765957447\\
0.59	1.00106382978723\\
0.6	1\\
0.61	1\\
0.62	1\\
0.63	1\\
0.64	1\\
0.65	1\\
0.66	1\\
0.67	1\\
0.68	1\\
0.69	1\\
0.7	1\\
0.71	1\\
0.72	1\\
0.73	1\\
0.74	1\\
0.75	1\\
0.76	1\\
0.77	1\\
0.78	1\\
0.79	1\\
0.8	1\\
0.81	1\\
0.82	1\\
0.83	1\\
0.84	1\\
0.85	1\\
0.86	1\\
0.87	1\\
0.88	1\\
0.89	1\\
0.9	1\\
0.91	1\\
0.92	1\\
0.93	1\\
0.94	1\\
0.95	1\\
0.96	1\\
0.97	1\\
0.98	1\\
0.99	1\\
1	1\\
};
\addlegendentry{{\Large $\delta_\mathrm{P,lb}(\mu,r_F,r_D)$ for $r_D=0$}}

\addplot [color=red, dotted, line width=2.0pt]
  table[row sep=crcr]{%
0	2.5\\
0.01	2.25\\
0.02	2\\
0.03	1.75\\
0.04	1.5\\
0.05	1.39285714285714\\
0.06	1.38571428571429\\
0.07	1.37857142857143\\
0.08	1.37142857142857\\
0.09	1.36428571428571\\
0.1	1.35714285714286\\
0.11	1.35\\
0.12	1.34285714285714\\
0.13	1.33571428571429\\
0.14	1.32857142857143\\
0.15	1.32142857142857\\
0.16	1.31428571428571\\
0.17	1.30714285714286\\
0.18	1.3\\
0.19	1.29285714285714\\
0.2	1.28571428571429\\
0.21	1.27857142857143\\
0.22	1.27142857142857\\
0.23	1.26428571428571\\
0.24	1.25714285714286\\
0.25	1.25\\
0.26	1.24285714285714\\
0.27	1.23571428571429\\
0.28	1.22857142857143\\
0.29	1.22142857142857\\
0.3	1.21428571428571\\
0.31	1.20714285714286\\
0.32	1.2\\
0.33	1.19285714285714\\
0.34	1.18571428571429\\
0.35	1.17857142857143\\
0.36	1.17142857142857\\
0.37	1.16428571428571\\
0.38	1.15714285714286\\
0.39	1.15\\
0.4	1.14285714285714\\
0.41	1.13571428571429\\
0.42	1.12857142857143\\
0.43	1.12142857142857\\
0.44	1.11428571428571\\
0.45	1.10714285714286\\
0.46	1.1\\
0.47	1.09285714285714\\
0.48	1.08571428571429\\
0.49	1.07857142857143\\
0.5	1.07142857142857\\
0.51	1.06428571428571\\
0.52	1.05714285714286\\
0.53	1.05\\
0.54	1.04285714285714\\
0.55	1.03571428571429\\
0.56	1.02857142857143\\
0.57	1.02142857142857\\
0.58	1.01428571428571\\
0.59	1.00714285714286\\
0.6	1\\
0.61	1\\
0.62	1\\
0.63	1\\
0.64	1\\
0.65	1\\
0.66	1\\
0.67	1\\
0.68	1\\
0.69	1\\
0.7	1\\
0.71	1\\
0.72	1\\
0.73	1\\
0.74	1\\
0.75	1\\
0.76	1\\
0.77	1\\
0.78	1\\
0.79	1\\
0.8	1\\
0.81	1\\
0.82	1\\
0.83	1\\
0.84	1\\
0.85	1\\
0.86	1\\
0.87	1\\
0.88	1\\
0.89	1\\
0.9	1\\
0.91	1\\
0.92	1\\
0.93	1\\
0.94	1\\
0.95	1\\
0.96	1\\
0.97	1\\
0.98	1\\
0.99	1\\
1	1\\
};
\addlegendentry{{\Large $\delta_\mathrm{P,ach}(\mu,r_F,r_D)$ for $r_D=0$}}

\addplot [color=mycolor1, dashdotted, line width=2.0pt]
  table[row sep=crcr]{%
0	2.5\\
0.01	2.25\\
0.02	2\\
0.03	1.75\\
0.04	1.5\\
0.05	1.25\\
0.06	1\\
0.07	1\\
0.08	1\\
0.09	1\\
0.1	1\\
0.11	1\\
0.12	1\\
0.13	1\\
0.14	1\\
0.15	1\\
0.16	1\\
0.17	1\\
0.18	1\\
0.19	1\\
0.2	1\\
0.21	1\\
0.22	1\\
0.23	1\\
0.24	1\\
0.25	1\\
0.26	1\\
0.27	1\\
0.28	1\\
0.29	1\\
0.3	1\\
0.31	1\\
0.32	1\\
0.33	1\\
0.34	1\\
0.35	1\\
0.36	1\\
0.37	1\\
0.38	1\\
0.39	1\\
0.4	1\\
0.41	1\\
0.42	1\\
0.43	1\\
0.44	1\\
0.45	1\\
0.46	1\\
0.47	1\\
0.48	1\\
0.49	1\\
0.5	1\\
0.51	1\\
0.52	1\\
0.53	1\\
0.54	1\\
0.55	1\\
0.56	1\\
0.57	1\\
0.58	1\\
0.59	1\\
0.6	1\\
0.61	1\\
0.62	1\\
0.63	1\\
0.64	1\\
0.65	1\\
0.66	1\\
0.67	1\\
0.68	1\\
0.69	1\\
0.7	1\\
0.71	1\\
0.72	1\\
0.73	1\\
0.74	1\\
0.75	1\\
0.76	1\\
0.77	1\\
0.78	1\\
0.79	1\\
0.8	1\\
0.81	1\\
0.82	1\\
0.83	1\\
0.84	1\\
0.85	1\\
0.86	1\\
0.87	1\\
0.88	1\\
0.89	1\\
0.9	1\\
0.91	1\\
0.92	1\\
0.93	1\\
0.94	1\\
0.95	1\\
0.96	1\\
0.97	1\\
0.98	1\\
0.99	1\\
1	1\\
};
\addlegendentry{{\Large $\delta_\mathrm{P}^*(\mu,r_F,r_D)$ for $r_D=0.6$}}

\end{axis}
\end{tikzpicture}%
	}
	\caption{Lower and upper bounds on the minimum NDT as a function of $\mu$ for $r_F=0.4$ and $M=K=10$.}
	\label{fig:pipe_ndt_vs_mu}
\end{figure}

The role of D2D cooperation in improving the delivery latency under pipelined delivery policies is further illustrated in \cref{fig:pipe_ndt_vs_mu}, where we plot the lower and upper bounds on the minimum NDT as a function of the fractional cache size $\mu$ for an F-RAN with $M=10$ ENs, $K=10$ users, and a fixed fronthaul rate $r_F=0.4$.
For small cache capacities satisfying $\mu\leq\mu_1$, D2D communication cannot reduce the minimum NDT because, in this regime, the total delivery time is dictated by fronthaul communication, which is required to deliver a large part of the requested files. In addition, for $\mu\geq 1-Mr_F/\min\{M,K\}$, 
the cache capacity is large enough to support delivery via cache-aided ZF with a fronthaul overhead that does not affect the achievability of the ideal NDT of one.
However, for $\mu_1<\mu<1-Mr_F/\min\{M,K\}$, a D2D-based scheme provides a latency reduction. For example, as depicted in \cref{fig:pipe_ndt_vs_mu}, for $r_D\geq 1-Mr_F/\min\{M,K\}$, an ideal NDT of one can be achieved with a fractional cache size $M$ times smaller than is required when no D2D communication is allowed ($r_D=0$).

\section{Conclusions}\label{sec:conclusions}
\looseness=-1
In this work, we have studied the benefits of out-of-band broadcast Device-to-Device (D2D) communication for content delivery in a general Fog-Radio Access Network (F-RAN) with arbitrary number of Edge Nodes (ENs) and users. Focusing on the normalized delivery time (NDT) metric, a strategy based on compress-and-forward D2D communication was shown to be approximately optimal to within a constant factor of $2$ for all values of the problem parameters, and under both serial and pipelined delivery policies. For sufficiently high D2D capacity, the proposed strategy was proved to achieve a significantly lower delivery latency than the minimum NDT for F-RAN without D2D communication.
Furthermore, we characterized the minimum NDT for the case of two ENs and users, and it was demonstrated that D2D communication can alleviate the load on the network infrastructure by reducing the traffic on the fronthaul links. 
Among related open problems we mention the design of robust delivery strategies that cope with the case in which some of the D2D links may be in outage; the case in which CSI at the ENs and cloud may be imperfect; the case in which inter-file coding is allowed; and the case in which security constraints are imposed on the ENs \cite{zeide2018confidential}.

\appendix 
\section{Appendices}

\subsection{Proof of Proposition \ref{prop:upper_bound}}\label{app:proof_ub_serial}
For the first three regimes, i.e., for low D2D rate $r_D\leq r_D^\text{th}$, the NDTs in \eqref{eq:ach_low_low_low}-\eqref{eq:ach_dc_high_low} are achieved by applying the strategy of \cite[Proposition 4]{sengupta2016fog}, which does not require D2D resources. 

Next, for low cache and high D2D rate, i.e., for $\mu\leq 1/M$ and $r_D>r_D^\text{th}$, a fraction $\mu M$ of each of the requested files is delivered via D2D-based CF, whereas the remaining $(1-\mu M)$ fraction is delivered via cloud-aided soft-transfer. 
The cache capacity constraint is satisfied since $\mu M\times 1/M+(1-\mu M)\times 0=\mu$, and the overall NDT is
\begin{IEEEeqnarray}{c}
	\delta_\text{ach}(\mu,r_F,r_D)=\mu M\delta_\text{D2D-CF}+(1-\mu M)\delta_\text{ST}.
\end{IEEEeqnarray}

\looseness=-1
Finally, for high cache and high D2D rate, i.e., for $\mu> 1/M$ and $r_D>r_D^\text{th}$, a fraction $(1-\mu)M/(M-1)$ of each of the requested files is delivered via D2D-based CF, whereas the remaining $(\mu M-1)/(M-1)$ fraction is delivered via cache-aided ZF.
The cache capacity constraint is satisfied since $(1-\mu) M/(M-1)\times(1/M)+(\mu M-1)/(M-1)\times 1=\mu$, and the overall NDT is
\begin{IEEEeqnarray}{c}
	\delta_\text{ach}(\mu,r_F,r_D)=\frac{(1-\mu)M}{M-1}\delta_\text{D2D-CF}+\frac{\mu M-1}{M-1}\delta_\text{ZF}.
\end{IEEEeqnarray}

\subsection{Proof of Proposition \ref{prop:lb}}\label{app:proof_lb}
For the proof of \cref{prop:lb}, we use the notation introduced in \cite[App. I]{sengupta2016fog}. Accordingly, for integers $0\leq a\leq b\leq K$ and $0\leq c\leq d\leq M$, we define $\mathbf{f}_{[a:b]}\triangleq(f_a,f_{a+1},\ldots,f_b)$, $\mathbf{s}_{[c:d]}\triangleq(s_c,s_{c+1},\ldots,s_d)$, $\mathcal{U}_{[c:d]}\triangleq \cb{\mathbf{u}_c,\mathbf{u}_{c+1},\ldots,\mathbf{u}_d}$, as well as the matrix of channel outputs
\begin{IEEEeqnarray}{c}
	\mathbf{Y}_{[a:b]}\triangleq\begin{bmatrix}
		y_a[1]&y_a[2]&\cdots&y_a[T_E]\\
		y_{a+1}[1]&y_{a+1}[2]&\cdots&y_{a+1}[T_E]\\
		\vdots&\vdots&\ddots&\vdots\\
		y_b[1]&y_b[2]&\cdots&y_b[T_E]
	\end{bmatrix},
\end{IEEEeqnarray}
and similarly for $\mathbf{Z}_{[a:b]}$ and $\mathbf{X}_{[c:d]}$. Furthermore, we define the following sub-matrix of the channel matrix $\mathbf{H}$
\begin{IEEEeqnarray}{c}
	\mathbf{H}_{[a:b]}^{[c:d]}\triangleq\begin{bmatrix}
		h_{a,c}&h_{a,c+1}&\cdots&h_{a,d}\\
		h_{a+1,c}&h_{a+1,c+1}&\cdots&h_{a+1,d}\\
		\vdots&\vdots&\ddots&\vdots\\
		h_{b,c}&h_{b,c+1}&\cdots&h_{b,d}
	\end{bmatrix}.
\end{IEEEeqnarray}

It follows from \cite[App. I]{sengupta2016fog} that, for $l=0,1,\ldots,\min\{M,K\}$,
\begin{IEEEeqnarray}{rCl}\label{eq:lb_eq1}
	KL&=&I\rb{\mathbf{f}_{[1:K]};\mathbf{Y}_{[1:l]},\mathcal{U}_{[1:(M-l)]},\mathbf{s}_{[1:(M-l)]}\given{\mathbf{f}_{[K+1:N]}}}\IEEEnonumber\\&&
	+H\rb{\mathbf{f}_{[1:K]}\given{\mathbf{Y}_{[1:l]},\mathcal{U}_{[1:(M-l)]},\mathbf{s}_{[1:(M-l)]},\mathbf{f}_{[K+1:N]}}},
\end{IEEEeqnarray}
and
\begin{IEEEeqnarray}{rCl}\label{eq:lb_eq2}
	I\rb{\mathbf{f}_{[1:K]};\mathbf{Y}_{[1:l]},\mathcal{U}_{[1:(M-l)]},\mathbf{s}_{[1:(M-l)]}\given{\mathbf{f}_{[K+1:N]}}}&\leq& H\rb{\mathbf{f}_{[1:l]}\given{\mathbf{Y}_{[1:l]}}}+(M-l)(K-l)\mu L\IEEEnonumber\\
	&&+lT_E\log(\Lambda P+1)+(M-l)r_FT_F\log(P),\IEEEeqnarraynumspace
\end{IEEEeqnarray}
where $\Lambda=\max_{k\in[l]}\sqb{\sum_{m=1}^{M}|h_{km}|^2+\sum_{m\neq\tilde{m}}h_{km}h_{k\tilde{m}}^*}$, and with the abuse of notation $\mathbf{Y}_{[1:0]}=\emptyset$ and $\mathbf{f}_{[1:0]}=\emptyset$.

We bound $H\rb{\mathbf{f}_{[1:l]}\given{\mathbf{Y}_{[1:l]}}}$ in \eqref{eq:lb_eq2} as follows
\begin{IEEEeqnarray}{rCl}
	H\rb{\mathbf{f}_{[1:l]}\given{\mathbf{Y}_{[1:l]}}}&=&H\rb{\mathbf{f}_{[1:l]}\given{\mathbf{Y}_{[1:l]},\mathcal V_1,\ldots,\mathcal V_l}}+I\rb{\mathbf{f}_{[1:l]};\mathcal V_1,\ldots,\mathcal V_l\given{\mathbf{Y}_{[1:l]}}}\IEEEnonumber\\
	&\leq&\sum_{k=1}^{l}H(f_k|\mathbf{y}_k,\mathcal V_k)+I\rb{\mathbf{f}_{[1:l]};\mathcal V_1,\ldots,\mathcal V_l\given{\mathbf{Y}_{[1:l]}}}\IEEEnonumber\\
	&\overset{\text{(a)}}{\leq}&lL\epsilon_L+I\rb{\mathbf{f}_{[1:l]};\mathcal V_1,\ldots,\mathcal V_l\given{\mathbf{Y}_{[1:l]}}}\IEEEnonumber\\
	&\leq&lL\epsilon_L+H\rb{\mathcal V_1,\ldots,\mathcal V_l},
\end{IEEEeqnarray}
where $\epsilon_L\geq 0$ is a function of $L$, independent of $P$, such that $\epsilon_L\rightarrow 0$ as $L\rightarrow\infty$; and (a) follows from Fano's inequality. For $l=0$, we have $\{\mathcal V_1,\ldots,\mathcal V_l\}=\emptyset$, whereas, for $l=1$, $\{\mathcal V_1,\ldots,\mathcal V_l\}=\{\mathbf{v}_2,\ldots,\mathbf{v}_K\}$, and, for $l=2$, $\{\mathcal V_1,\ldots,\mathcal V_l\}=\{\mathbf{v}_1,\ldots,\mathbf{v}_K\}$. Hence, $H\rb{\mathcal V_1,\ldots,\mathcal V_l}\leq g(l)T_Dr_D\log(P)$, where $g(l)$ is defined in~\eqref{eq:g_l}, and we can further bound $H\rb{\mathbf{f}_{[1:l]}\given{\mathbf{Y}_{[1:l]}}}$ as
\begin{IEEEeqnarray}{c}\label{eq:lb_eq3}
	H\rb{\mathbf{f}_{[1:l]}\given{\mathbf{Y}_{[1:l]}}}\leq lL\epsilon_L+g(l)T_Dr_D\log(P).
\end{IEEEeqnarray}

Next, we bound $H\rb{\mathbf{f}_{[1:K]}\given{\mathbf{Y}_{[1:l]},\mathcal{U}_{[1:(M-l)]},\mathbf{s}_{[1:(M-l)]},\mathbf{f}_{[K+1:N]}}}$ in \eqref{eq:lb_eq1} as follows
\begin{IEEEeqnarray}{l}
	H\rb{\mathbf{f}_{[1:K]}\given{\mathbf{Y}_{[1:l]},\mathcal{U}_{[1:(M-l)]},\mathbf{s}_{[1:(M-l)]},\mathbf{f}_{[K+1:N]}}}\IEEEnonumber\\
	\qquad\qquad=H\rb{\mathbf{f}_{[1:K]}\given{\mathbf{Y}_{[1:l]},\mathbf{Y}_{[l+1:K]},\mathcal{U}_{[1:(M-l)]},\mathbf{s}_{[1:(M-l)]},\mathbf{f}_{[K+1:N]}}}\IEEEnonumber\\
	\qquad\qquad\quad+I\rb{\mathbf{f}_{[1:K]};\mathbf{Y}_{[l+1:K]}\given{\mathbf{Y}_{[1:l]},\mathcal{U}_{[1:(M-l)]},\mathbf{s}_{[1:(M-l)]},\mathbf{f}_{[K+1:N]}}}\IEEEnonumber\\
	\qquad\qquad\leq H\rb{\mathbf{f}_{[1:K]}\given{\mathbf{Y}_{[1:K]}}}	
	+H\rb{\mathbf{Y}_{[l+1:K]}\given{\mathbf{Y}_{[1:l]},\mathcal{U}_{[1:(M-l)]},\mathbf{s}_{[1:(M-l)]},\mathbf{f}_{[K+1:N]}}}\IEEEnonumber\\
	\qquad\qquad\quad-H\rb{\mathbf{Y}_{[l+1:K]}\given{\mathbf{Y}_{[1:l]},\mathcal{U}_{[1:(M-l)]},\mathbf{s}_{[1:(M-l)]},\mathbf{f}_{[1:N]}}}\IEEEnonumber\\
	\qquad\qquad\leq KL\epsilon_L+H\rb{\mathbf{Y}_{[l+1:K]}\given{\mathbf{Y}_{[1:l]},\mathbf{X}_{[1:(M-l)]}}}-H\rb{\mathbf{Z}_{[l+1:K]}}.
\end{IEEEeqnarray}
By applying \cite[Lemma 7]{sengupta2016fog}, we get
\begin{IEEEeqnarray}{rCl}
	H\rb{\mathbf{Y}_{[l+1:K]}\given{\mathbf{Y}_{[1:l]},\mathbf{X}_{[1:(M-l)]}}}&=&H\rb{\mathbf{Y}_{[l+1:K]}\given{\mathbf{Y}_{[1:l]},\mathbf{X}_{[1:(M-l)]},\mathbf{Y}_{[l+1:K]}+\tilde{\mathbf{Z}}_{[l+1:K]}-\mathbf{Z}_{[l+1:K]}}}\IEEEnonumber\\
	&\leq&H\rb{\mathbf{Y}_{[l+1:K]}\given{\mathbf{Y}_{[l+1:K]}+\tilde{\mathbf{Z}}_{[l+1:K]}-\mathbf{Z}_{[l+1:K]}}}\IEEEnonumber\\
	&\leq&H\rb{\tilde{\mathbf{Z}}_{[l+1:K]}-\mathbf{Z}_{[l+1:K]}},
\end{IEEEeqnarray}
where we define $\tilde{\mathbf{Z}}_{[l+1:K]}\triangleq(\mathbf{H}_2\cdot\mathbf{H}_1^\dagger)\mathbf{Z}_{[1:l]}$ with $\mathbf{H}_1\triangleq\mathbf{H}_{[1:l]}^{[(M-l)+1:M]}$ and $\mathbf{H}_2\triangleq\mathbf{H}_{[l+1:K]}^{[(M-l)+1:M]}$. Matrix $\mathbf{H}_1^\dagger$ is the Moore-Penrose pseudo-inverse of $\mathbf{H}_1$. Therefore,
\begin{IEEEeqnarray}{rCl}\label{eq:lb_eq4}
	H\rb{\mathbf{f}_{[1:K]}\given{\mathbf{Y}_{[1:l]},\mathcal{U}_{[1:(M-l)]},\mathbf{s}_{[1:(M-l)]},\mathbf{f}_{[K+1:N]}}}&\leq&KL\epsilon_L+T_E\log\det\rb{\mathbf{I}_{[K-l]}+\tilde{\mathbf{H}}\tilde{\mathbf{H}}^H},\IEEEeqnarraynumspace
\end{IEEEeqnarray}
where $\tilde{\mathbf{H}}\triangleq\mathbf{H}_2\cdot\mathbf{H}_1^\dagger$.

Overall, for $l=0,1,\ldots,\min\{M,K\}$, it follows from \eqref{eq:lb_eq1}, \eqref{eq:lb_eq2}, \eqref{eq:lb_eq3}, and~\eqref{eq:lb_eq4} that
\begin{IEEEeqnarray}{rCl}
	K&\leq& (M-l)(K-l)\mu +l\frac{T_E}{L}\log(\Lambda P+1)+(M-l)r_F\frac{T_F}{L}\log(P)\IEEEnonumber\\
	&&+(K+l)\epsilon_L+\frac{T_E}{L}\log\det\rb{\mathbf{I}_{[K-l]}+\tilde{\mathbf{H}}\tilde{\mathbf{H}}^H}+g(l)\frac{T_D}{L}r_D\log(P).
\end{IEEEeqnarray}
Now, we take the limit $L\rightarrow\infty$ and then $P\rightarrow\infty$, and arrive at \eqref{eq:lb_family}.

Finally, since the Degrees of Freedom (DoF) of \eqref{eq:wireless_channel} are upper bounded by the DoF of the $M\times K$ MIMO point-to-point channel, i.e., $\min\{M,K\}$ \cite{telatar1999capacity}, then
\begin{IEEEeqnarray}{l}
	\delta_E=\lim_{P\rightarrow\infty}\lim_{L\rightarrow\infty}\frac{T_E}{L/\log(P)}
	\geq \lim_{P\rightarrow\infty}\lim_{L\rightarrow\infty}\frac{KL/(\min\{M,K\}\log(P))}{L/\log(P)}\
	=\frac{K}{\min\{M,K\}},\IEEEeqnarraynumspace
\end{IEEEeqnarray} 
i.e., inequality \eqref{eq:lb_DoF}. 
Inequalities \eqref{eq:lb_trivial} follows trivially from the definitions of $\delta_F$ and $\delta_D$ \eqref{eq:F_NDT}.

\subsection{Proof of Proposition \ref{prop:min_ndt_serial_2x2}}\label{app:proof_min_ndt_serial_2x2}
Since the achievability was established by \cref{prop:ub_2x2}, here we prove the converse, i.e., $\delta^*(\mu,r_F,r_D)\geq\delta_{2\times2}(\mu,r_F,r_D)$.
For $M=K=2$, the constraints~\eqref{eq:lb_family},~\eqref{eq:lb_DoF}, in \cref{prop:lb} can be written as:
\begin{IEEEeqnarray}{rCl}
	\delta_E+r_F\delta_F+r_D\delta_D&\geq&2-\mu,\label{eq:lb_2x2_eq1}\\
	\delta_F&\geq&\frac{1-2\mu}{r_F},\label{eq:lb_2x2_eq2}\\
	\delta_E&\geq&1.\label{eq:lb_2x2_eq3}
\end{IEEEeqnarray}

For $0\leq r_F,r_D\leq 1$, using \eqref{eq:lb_2x2_eq1} gives the lower bound $\delta^*(\mu,r_F,r_D)\geq 2-\mu$. Furthermore, using $(1-r_F)\times\eqref{eq:lb_2x2_eq2}+\eqref{eq:lb_2x2_eq1}$ gives the lower bound $\delta^*(\mu,r_F,r_D)\geq 1+\mu+(1-2\mu)/r_F$.

For $r_F\geq\max\{1,r_D\}$, using $[(r_F-1)\times\eqref{eq:lb_2x2_eq3}+\eqref{eq:lb_2x2_eq1}]/r_F$ gives the lower bound $\delta^*(\mu,r_F,r_D)\geq 1+(1-\mu)/r_F$.

For $r_D>\max\{1,r_F\}$, using $[(r_D-1)\times\eqref{eq:lb_2x2_eq3}+\eqref{eq:lb_2x2_eq1}]/r_D$ gives the lower bound $\delta^*(\mu,r_F,r_D)\geq 1+(1-\mu)/r_D$. Moreover, using $[\eqref{eq:lb_2x2_eq1}+(r_D-r_F)\times\eqref{eq:lb_2x2_eq2}+(r_D-1)\times\eqref{eq:lb_2x2_eq3}]/r_D$ gives the lower bound $\delta^*(\mu,r_F,r_D)\geq 1+\mu/r_D+(1-2\mu)/r_F$.

\subsection{Proof of Proposition \ref{prop:char}}\label{app:proof_ratio2}
We prove \cref{prop:char} by showing that the ratio \eqref{eq:NDT_ratio} holds in each of the five regimes described in \cref{prop:upper_bound}. First, note that, due to \eqref{eq:lb_DoF}-\eqref{eq:lb_trivial}, we have
\begin{IEEEeqnarray}{c}\label{eq:app_DoF_lb}
	\delta^*(\mu,r_F,r_D)\geq \frac{K}{\min\{M,K\}}.
\end{IEEEeqnarray}
Another lower bound on the minimum NDT follows from \eqref{eq:lb_family}-\eqref{eq:lb_trivial} (with $l=0$) as
\begin{IEEEeqnarray}{c}\label{eq:app_lb2}
	\delta^*(\mu,r_F,r_D)\geq \frac{K}{\min\{M,K\}}+\frac{K\rb{1-M\mu}}{Mr_F}.
\end{IEEEeqnarray}

\subsubsection{High Cache and High D2D ($\mu> 1/M$ and $r_D> r_D^\text{th}$)}
Since $\mu> 1/M$, the achievable NDT \eqref{eq:ach_high_dc_high} satisfies
\begin{IEEEeqnarray}{c}\label{eq:app_ach_tmp1}
	\delta_\text{ach}(\mu,r_F,r_D)\leq\frac{K}{\min\{M,K\}}\rb{1+\frac{1}{r_D}}.
\end{IEEEeqnarray}
Dividing~\eqref{eq:app_ach_tmp1} by~\eqref{eq:app_DoF_lb} gives
\begin{IEEEeqnarray}{c}\label{eq:1/rD_1}
	\frac{\delta_\text{ach}(\mu,r_F,r_D)}{\delta^*(\mu,r_F,r_D)}\leq 1+\frac{1}{r_D}.
\end{IEEEeqnarray}
Next, since $r_D>r_D^\text{th}$, where $r_D^\text{th}$ is the threshold defined in~\eqref{eq:D2D_th}, we have
\begin{IEEEeqnarray}{c}\label{eq:rD_g_1}
	r_D>r_D^\text{th}\geq \frac{\max\{M,K\}}{\min\{M,K\}-1}>1.
\end{IEEEeqnarray}
Thus, 
\begin{IEEEeqnarray}{c}
	\frac{\delta_\text{ach}(\mu,r_F,r_D)}{\delta^*(\mu,r_F,r_D)}<2.
\end{IEEEeqnarray}

\subsubsection{Low Cache and High D2D ($\mu\leq 1/M$ and $r_D> r_D^\text{th}$)}
Dividing~\eqref{eq:ach_low_dc_high} by~\eqref{eq:app_lb2} gives
\begin{IEEEeqnarray}{rCl}\label{eq:1/rD_2}
	\frac{\delta_\text{ach}(\mu,r_F,r_D)}{\delta^*(\mu,r_F,r_D)}&\leq&1+\frac{\max\{M,K\}\mu/r_D}{K/\min\{M,K\}+K(1-\mu M)/(Mr_F)}\IEEEnonumber\\
	&\overset{\text{(a)}}{\leq}&1+\frac{1}{r_D}
	\overset{\text{(b)}}{<}2,	
\end{IEEEeqnarray}
where (a) follows from $\mu\leq 1/M$, and (b) is due to $r_D>1$ \eqref{eq:rD_g_1}.

\subsubsection{High Cache, Low Fronthaul, and Low D2D ($\mu> 1/M$, $r_F\leq r_F^\text{th}$, and $r_D\leq r_D^\text{th}$)}
In this regime, we have
\begin{IEEEeqnarray}{rCl}\label{eq:app_achievable_tmp2}
	\delta_\text{ach}(\mu,r_F,r_D)&=&\frac{K}{\min\{M,K\}}\rb{\frac{\mu M-1}{M-1}}+(1-\mu)\frac{M+K-1}{M-1}\IEEEnonumber\\
	&\leq& \frac{M-1}{M}\cdot\frac{M+K-1}{M-1}
	=\frac{M+K-1}{M},
\end{IEEEeqnarray}
where the inequality follows from $\delta_\text{ach}(\mu,r_F,r_D)$ being a monotonically decreasing function of $\mu$ and since $\mu\geq 1/M$. 
Dividing~\eqref{eq:app_achievable_tmp2} by~\eqref{eq:app_DoF_lb} gives
\begin{IEEEeqnarray}{rCl}
	\frac{\delta_\text{ach}(\mu,r_F,r_D)}{\delta^*(\mu,r_F,r_D)}\leq\frac{M+K-1}{M}\cdot\frac{\min\{M,K\}}{K}
	&=&1+\frac{\min\{M,K\}-1}{\max\{M,K\}}
	<2.
\end{IEEEeqnarray}

\subsubsection{High Fronthaul and Low D2D ($r_F> r_F^\text{th}$ and $r_D\leq r_D^\text{th}$)}
In this regime, we have 
\begin{IEEEeqnarray}{rCl}\label{eq:app_ach_tmp3}
	\delta_\text{ach}(\mu,r_F,r_D)&=&\frac{K}{\min\{M,K\}}+\frac{(1-\mu)K}{Mr_F}\IEEEnonumber\\
	&\overset{\text{(a)}}{\leq}&\frac{K}{\min\{M,K\}}+\frac{K}{Mr_F}\IEEEnonumber\\
	&\overset{\text{(b)}}{<}&\frac{K}{\min\{M,K\}}+\frac{\min\{M,K\}-1}{M-1},
\end{IEEEeqnarray}
where (a) follows from $\mu\geq0$, and (b) follows from $r_F>r_F^\text{th}$.
Dividing~\eqref{eq:app_ach_tmp3} by~\eqref{eq:app_DoF_lb} gives
\begin{IEEEeqnarray}{rCl}
	\frac{\delta_\text{ach}(\mu,r_F,r_D)}{\delta^*(\mu,r_F,r_D)}&<&1+\frac{\min\{M,K\}-1}{M-1}\cdot\frac{\min\{M,K\}}{K}\leq 2.
\end{IEEEeqnarray} 

\subsubsection{Low Cache, Low Fronthaul, and Low D2D ($\mu\leq 1/M$, $r_F\leq r_F^\text{th}$, and $r_D\leq r_D^\text{th}$)}
We first consider the case of $K\leq M$. 
Dividing~\eqref{eq:ach_low_low_low} by~\eqref{eq:app_lb2} gives
\begin{IEEEeqnarray}{rCl}
	\frac{\delta_\text{ach}(\mu,r_F,r_D)}{\delta^*(\mu,r_F,r_D)}&\leq&1+\frac{\mu(K-1)}{1+K(1-M\mu)/(Mr_F)}\IEEEnonumber\\
	&\overset{\text{(a)}}{\leq}&1+\frac{K-1}{M}
	<2,	
\end{IEEEeqnarray}
where (a) follows from $\mu\leq 1/M$.

Next, for $M<K$ and $r_F\geq 1$, the achievable NDT \eqref{eq:ach_low_low_low} satisfies
\begin{IEEEeqnarray}{rCl}\label{eq:app_ach_tmp5}
	\delta_\text{ach}(\mu,r_F,r_D)&\overset{\text{(a)}}{\leq}&\frac{K}{M}+\frac{K}{Mr_F}
	\overset{\text{(b)}}{\leq}\frac{2K}{M},
\end{IEEEeqnarray}
where (a) follows from $\delta_\text{ach}(\mu,r_F,r_D)$ being a monotonically decreasing function of $\mu$ and since $\mu>0$, whereas (b) is due to $r_F\geq 1$. 
Dividing~\eqref{eq:app_ach_tmp5} by~\eqref{eq:app_DoF_lb} gives
\begin{IEEEeqnarray}{c}
	\frac{\delta_\text{ach}(\mu,r_F,r_D)}{\delta^*(\mu,r_F,r_D)}\leq \frac{2K}{M}\cdot\frac{M}{K}=2.
\end{IEEEeqnarray}

\looseness=-1
Finally, we consider the case of $M<K$ and $r_F<1$. 
Let the integer $l^*$ be defined as $l^*\triangleq\lceil M/2\rceil$.
Note that since we consider a case with $r_F<1$, then we have the following inequality
\begin{IEEEeqnarray}{c}
	(M-l^*)r_F\leq l^*.\label{eq:tmp6}
\end{IEEEeqnarray}
We further divide the case of $M<K$ and $r_F<1$ into two regimes: $r_D\leq l^*/g(l^*)$ and $l^*/g(l^*)<r_D\leq r_D^\text{th}$. For $r_D\leq l^*/g(l^*)$, it follows from \eqref{eq:lb_family} with $l=l^*$ that 
\begin{IEEEeqnarray}{c}\label{eq:tmp7}
	l^*\delta_E+(M-l^*)r_F\delta_F+l^*\delta_D\geq K-(M-l^*)(K-l^*)\mu.
\end{IEEEeqnarray} 
Furthermore, we have
\begin{IEEEeqnarray}{c}\label{eq:tmp8}
	[l^*-(M-l^*)r_F]\delta_F\geq[l^*-(M-l^*)r_F]\cdot\frac{K(1-M\mu)}{Mr_F}
\end{IEEEeqnarray}
due to \eqref{eq:lb_family} (with $l=0$) and \eqref{eq:tmp6}. By adding \eqref{eq:tmp7} and \eqref{eq:tmp8}, and dividing by $l^*$, we get the following lower bound on the minimum NDT
\begin{IEEEeqnarray}{rCl}\label{eq:app_lb_serial_big_case}
	\delta^*(\mu,r_F,r_D)&\geq&
	\frac{K(1-M\mu)}{M}\rb{1+\frac{1}{r_F}}+(M+K-l^*)\mu.
\end{IEEEeqnarray}
Dividing~\eqref{eq:ach_low_low_low} by~\eqref{eq:app_lb_serial_big_case} gives
\begin{IEEEeqnarray}{rCl}
	\frac{\delta_\text{ach}(\mu,r_F,r_D)}{\delta^*(\mu,r_F,r_D)}&\leq&1+\frac{(l^*-1)\mu}{K(1-M\mu)\cdot(1+1/r_F)/M+(M+K-l^*)\mu}\IEEEnonumber\\
	&\leq&1+\frac{l^*-1}{M+K-l^*}\IEEEnonumber\\
	&\overset{\text{(a)}}{<}&1+\frac{M/2}{M/2+K-1}
	<2,
\end{IEEEeqnarray}
where (a) follows from $l^*< M/2+1$. 

Now, for $l^*/g(l^*)<r_D\leq r_D^\text{th}$, we have $g(l^*)r_D>l^*\geq (M-l^*)r_F$. 
Thus, \eqref{eq:lb_family} (for $l=0$) and \eqref{eq:lb_DoF} imply, respectively,
\begin{IEEEeqnarray}{rCl}
	(g(l^*)r_D-(M-l^*)r_F)\delta_F &\geq&(g(l^*)r_D-(M-l^*)r_F)\cdot \frac{K(1-M\mu)}{Mr_F}, \label{eq:tmp10}\\
	(g(l^*)r_D-l^*)\delta_E&\geq& (g(l^*)r_D-l^*)\cdot\frac{K}{M}.\label{eq:tmp11}
\end{IEEEeqnarray}
By adding \eqref{eq:tmp10} and \eqref{eq:tmp11} to \eqref{eq:lb_family} (with $l=l^*$), and dividing by $g(l^*)r_D$, we get the following lower bound on the minimum NDT
\begin{IEEEeqnarray}{rCl}\label{eq:app_tmp_lb}
	\delta^*(\mu,r_F,r_D)&\geq&
	\frac{K(1-M\mu)}{M}\rb{1+\frac{1}{r_F}}+K\mu+\frac{(M-l^*)l^*\mu}{g(l^*)r_D}\IEEEnonumber\\
	&\overset{\text{(a)}}\geq&\frac{K(1-M\mu)}{M}\rb{1+\frac{1}{r_F}}+K\mu+\frac{(M-l^*)(M-1)l^*\mu}{g(l^*)K},
\end{IEEEeqnarray}
where (a) follows from $r_D\leq r_D^\text{th}= K/(M-1)$. 
Dividing~\eqref{eq:ach_low_low_low} by~\eqref{eq:app_tmp_lb} gives
\begin{IEEEeqnarray}{rCl}
	\frac{\delta_\text{ach}(\mu,r_F,r_D)}{\delta^*(\mu,r_F,r_D)}&\leq&1+\frac{(M-1)\mu[1-(M-l^*)l^*/(g(l^*)K)]}{K(1-M\mu)\cdot(1+1/r_F)/M+K\mu+(M-l^*)(M-1)l^*\mu/(g(l^*)K)}\IEEEnonumber\\
	&\leq&1+\frac{(M-1)\mu}{K\mu}
	<2.
\end{IEEEeqnarray}

\subsection{Proof of Proposition \ref{prop:upper_bound_pipe}}\label{app:proof_ub_pipe}
For high fronthaul rate, $r_F\geq \min\{M,K\}/M$, we apply block-Markov encoding with cloud-aided soft-transfer \cite[Proposition 3]{sengupta2016fog}; the resulting NDT is
\begin{IEEEeqnarray}{rCl}
	\delta_\text{P}\rb{\mu,r_F,r_D}&=&\max\cb{\frac{K}{Mr_F},\frac{K}{\min\{M,K\}},0}=\frac{K}{\min\{M,K\}}.
\end{IEEEeqnarray}
Note that, in this regime, no caching and D2D resources are required.

Next, we consider low fronthaul rate, i.e., $r_F< \min\{M,K\}/M$.
For $\mu\in[0,\mu_1]$, where $\mu_1$ is defined in \eqref{eq:pipe_mu1}, no D2D communication is utilized. As in \cite[Proposition 9]{sengupta2016fog}, we apply the following per-block file-splitting with block-Markov encoding: Part $(1-M\mu)$ of each requested file is delivered using cloud-aided soft-transfer \cite[Proposition 3]{sengupta2016fog}; and part $M\mu$ of each requested file is delivered using cache-aided EN coordination \cite[Lemma 3]{sengupta2016fog}. The cache capacity constraint is satisfied since $\mu\leq\mu_1\leq 1/M$. This achieves the NDT
\begin{IEEEeqnarray}{rCl}\label{eq:pipe_proof_ach_ndt_1}
	\delta_\text{P}\rb{\mu,r_F,r_D}&=&\max\cb{\frac{(1-M\mu)K}{Mr_F},\frac{(1-M\mu)K}{\min\{M,K\}}+\frac{M\mu(M+K-1)}{M},0}
	=\frac{(1-M\mu)K}{Mr_F},\IEEEeqnarraynumspace
\end{IEEEeqnarray}
where the last equality follows from $\mu\leq\mu_1$.

For $\mu\in[\mu_2,1]$, where $\mu_2$ is defined in \eqref{eq:pipe_mu2}, we apply the following per-block file-splitting with block-Markov encoding: Part $\alpha_1\triangleq \min\{Mr_F/\min\{M,K\},1\}$ of each requested file is delivered using cloud-aided soft-transfer; part $\alpha_2\triangleq\min\{r_D,1-\alpha_1\}$ of each requested file is delivered using D2D-based compress-and-forward (\cref{prop:CF}); and part $(1-\alpha_1-\alpha_2)$ of each requested file is delivered using cache-aided ZF \cite[Lemma 2]{sengupta2016fog}. The cache capacity constraint is satisfied since $1-\alpha_1-\alpha_2+\alpha_2/M=\mu_2\leq\mu$. This achieves the NDT
\begin{IEEEeqnarray}{rCl}
	\delta_\text{P}\rb{\mu,r_F,r_D}&=&\max\cb{\alpha_1\cdot\frac{K}{Mr_F},\frac{K}{\min\{M,K\}},\alpha_2\cdot\frac{K}{r_D\min\{M,K\}}}
	=\frac{K}{\min\{M,K\}}.
\end{IEEEeqnarray}  

Finally, for $\mu\in[\mu_1,\mu_2]$, we apply file-splitting and cache-sharing \cite[Lemma 1]{sengupta2016fog} between the policies for the corner points $\mu=\mu_1$ and $\mu=\mu_2$. This achieves the NDT
\begin{IEEEeqnarray}{rCl}
	\delta_\text{P}\rb{\mu,r_F,r_D}&=&\frac{\mu_2-\mu}{\mu_2-\mu_1}\cdot\frac{(1-M\mu_1)K}{Mr_F}+\frac{\mu-\mu_1}{\mu_2-\mu_1}\cdot\frac{K}{\min\{M,K\}}.
\end{IEEEeqnarray} 

\subsection{Proof of Proposition \ref{prop:pipe_ndt_ratio}}\label{app:pipe_ndt_ratio}
In \cite[App. VIII-C]{sengupta2016fog}, it was proved that, without D2D communication,
\begin{IEEEeqnarray}{c}\label{eq:pipe_no_d2d_ratio}
	\frac{\delta_\text{P,ach}(\mu,r_F,r_D=0)}{K/\min\{M,K\}}\leq 2,\quad\forall\mu\in[\mu_1,\tilde{\mu}_2], 
\end{IEEEeqnarray}
where $\tilde{\mu}_2\triangleq1-Mr_F/\min\{M,K\}$. Thus, for all $\mu\in[\mu_1,\mu_2]$, the ratio between the achievable NDT and the minimum NDT under pipelined delivery is upper bounded as
\begin{IEEEeqnarray}{rCL}
	\frac{\delta_\text{P,ach}(\mu,r_F,r_D)}{\delta_\text{P}^*(\mu,r_F,r_D)}&\overset{\text{(a)}}{\leq}&\frac{\delta_\text{P,ach}(\mu,r_F,r_D)}{K/\min\{M,K\}}
	\overset{\text{(b)}}{\leq}\frac{\delta_\text{P,ach}(\mu,r_F,r_D=0)}{K/\min\{M,K\}}
	\overset{\text{(c)}}{\leq}2,
\end{IEEEeqnarray} 
where (a) follows from \cref{cor:piple_lb}; (b) holds since D2D cooperation does not increase the achievable NDT of \cref{prop:upper_bound_pipe}; and (c) follows from~\eqref{eq:pipe_no_d2d_ratio} and since $\tilde{\mu}_2\geq\mu_2$.

\subsection{Proof of Proposition \ref{prop:pipe_min_NDT}}\label{app:proof_pipe_min_ndt}
The lower bound of \cref{cor:piple_lb} can be relaxed by considering only $l=0$, i.e.,
\begin{IEEEeqnarray}{c}\label{eq:pipe_lb_l=0}
	\delta_\text{P}^*\rb{\mu,r_F,r_D}\geq\max\cb{\frac{(1-M\mu)K}{Mr_F},\frac{K}{\min\{M,K\}}}.
\end{IEEEeqnarray}
For high fronthaul rate and for low fronthaul rate with fractional cache capacity $\mu$ that satisfies $\mu\in[0,\mu_1]$ or $\mu\in[\mu_2,1]$, the lower bound~\eqref{eq:pipe_lb_l=0} coincides with the achievable NDT of \cref{prop:upper_bound_pipe} (\eqref{eq:pipe_ach_ndt1} and~\eqref{eq:pipe_ach_ndt2}), hence the minimum NDT is given by~\eqref{eq:pipe_min_NDT1} and~\eqref{eq:pipe_min_NDT2}, respectively. Next, for low fronthaul rate $r_F<\min\{M,K\}/M$ and high D2D rate $r_D\geq 1-Mr_F/\min\{M,K\}$, the strategy of \cref{prop:upper_bound_pipe} achieves an NDT of $\delta_\text{P,ach}(\mu,r_F,r_D)=K/\min\{M,K\}$ for all $\mu\geq \mu_2=1/M-r_F/\min\{M,K\}$; and an NDT of $\delta_\text{P,ach}(\mu,r_F,r_D)=K/(Mr_F)$ for cloud-only F-RAN, i.e., for $\mu=0$ (see~\eqref{eq:pipe_ach_ndt2}). For $\mu\in(0,\mu_2)$ we apply file-splitting and cache-sharing between the policies for $\mu=0$ and $\mu=\mu_2$.
This achieves the NDT~\eqref{eq:pipe_min_ndt_high_rd}, which equals the lower bound~\eqref{eq:pipe_lb_l=0}, and hence optimal.

\bibliographystyle{IEEEtranTCOM}
\bibliography{IEEEabrv,../../../../TexInputs/myBib}

\end{document}